\documentclass[preprints,review,accept,pdftex]{Definitions/mdpi} 

\usepackage{amssymb} 
\usepackage[normalem]{ulem}
\usepackage{framed}
\usepackage{mathtools}
\usepackage[T1]{fontenc}
\usepackage[utf8]{inputenc}
\newcommand{\be}{\begin{equation}}
\newcommand{\ee}{\end{equation}}
\newcommand{\bal}{\begin{align}}
\newcommand{\eal}{\end{align}}

\newcommand{\hf}{\hat{ \varphi}}
\newcommand{\hvf}{\hat{\bm \varphi}}

\newcommand{\apj}{Astrophys. J.}

 \def\Tr{\textnormal{Tr}}
  
 \def\({\left(}
 \def\){\right)}
 \def\[{\left[}
 \def\]{\right]}

\usepackage{bm}

\newcommand{\beq}{\begin{equation}}
\newcommand{\eeq}{\end{equation}}
\newcommand{\bea}{\begin{eqnarray}}
\newcommand{\eea}{\end{eqnarray}}

\setitemize{parsep=6pt,itemsep=0pt,leftmargin=*,labelsep=5.5mm}
\setenumerate{parsep=6pt,itemsep=0pt,leftmargin=*,labelsep=5.5mm}
\setlist[description]{itemsep=0mm}   

\firstpage{1} 
\pubvolume{xx}
\issuenum{1}
\articlenumber{5}
\pubyear{2019}
\copyrightyear{2019}
\history{Received: date; Accepted: date; Published: date}
\updates{yes} 

\Title{Meson Condensation}

\Author{Massimo  Mannarelli}

\AuthorNames{Massimo  Mannarelli}

\address{%
INFN-Laboratori Nazionali del Gran Sasso, Via G. Acitelli, 22, I-67100 Assergi (AQ), Italy;\\ 	{massimo.mannarelli}@lngs.infn.it}

\corres{massimo.mannarelli@lngs.infn.it}

\firstnote{} 

\abstract{ We give a pedagogical review of the properties of the various meson condensation phases   triggered by a large isospin or strangeness imbalance. We argue that these phases are  extremely interesting and powerful playground for exploring the properties of hadronic matter. The reason is that they are realized in a regime in which various theoretical methods overlap with increasingly precise numerical lattice QCD simulations, providing insight on the properties of color confinement and of chiral symmetry breaking. }


\begin{document}


\section{Introduction}\label{sec:Introduction}

The great success of the Standard Model of particle physics relies on the possibility of making accurate and testable predictions that are in agreement with increasingly precise experimental data. Despite this success, many aspects of the Standard Model are still not completely clear. Among these,  there are the mechanisms of color confinement  and of chiral symmetry breaking ($\chi$SB) of the strong interaction. The    typical energy scales of  confinement  and $\chi$SB pertain to the  nonperturbative region of quantum chromodynamics (QCD), which makes their study extremely challenging.

For some theoreticians,  the   confinement and  the $\chi$SB  mechanisms are uninteresting because they are {\it details} of a robust theoretical construction, thus   sooner or later they will be fully understood; for~ others (including the author), unraveling  the origin of these  mechanisms is of the utmost importance for a comprehensive understanding of QCD; for all, it is still unclear which is the path that can bring us to a full understanding of these mechanisms. My view is that any path, as far it is physically sound, should be explored and tested. This brief review is about one those paths, exploring the behavior of matter when there is an asymmetry in the number of particles with different isospin and/or  strangeness. This seems a promising direction because a number of theoretical methods can be used for studying these phases. Comparing the results obtained by different methods helps us to check their consistency   and the degree of the reached accuracy. 

\begin{figure}[H]
\centering
\includegraphics[width=.7\textwidth]{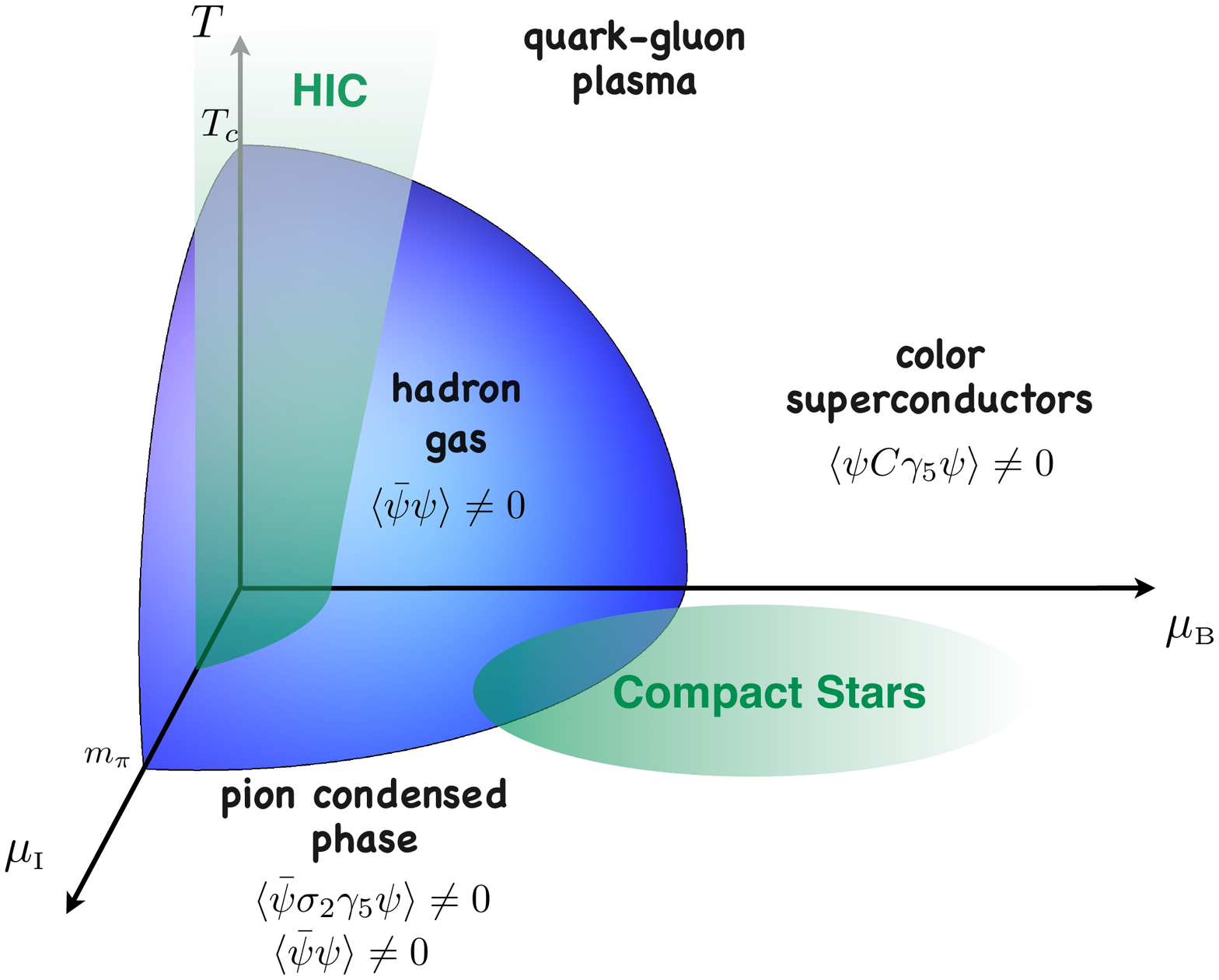}
\hspace*{-2.cm}
\caption{Cartoon of the grand-canonical phase diagram of hadronic matter as a function of temperature and of the isospin  and  baryonic chemical potentials. The  shaded green regions   are explored by heavy ion collisions (HIC) or   are possibly realized in compact stars. For each phase we have indicated the relevant quark condensates, {see Equations~(\ref{eq:chiral_cond})--(\ref{eq:diquark_cond}) below}.    }
\label{fig:phase_diagram}
 \end{figure}

To clarify the setting we report in Figure~\ref{fig:phase_diagram}  a sketch of the so-called QCD phase diagram: a grand canonical description of the phases of hadronic matter as a function of the hadronic temperature, $T$, of the isospin chemical potential, $\mu_I$, and of the baryonic chemical potentials, $\mu_B$. The total baryonic density is determined by  $\mu_B$, while $\mu_I$ describes the isospin asymmetry,  say due to a different number of up and down quarks. If it were possible  we would have added a further axis, $\mu_S$, indicating the strangeness content. 
The blue region corresponds to a gas of confined hadrons with a chiral broken symmetry. At large energy scales quarks and gluons should be liberated~\cite{Cabibbo:1975ig} realizing different phases. We have indicated three of them. The quark-gluon plasma (QGP), realized at large temperature,  is~ asymptotically a gas of quarks and gluons that becomes strongly interacting for the temperature reachable in heavy-ion collisions, see for instance~\cite{Gyulassy:2004vg,Shuryak:2008eq,Satz:2012zza}.  At large $\mu_B$ we expect that deconfined quarks fill their Fermi spheres and that the color interaction drives the  formation of Cooper pairs  in a BCS-like color superconducting phase  (CSC), see \cite{Rajagopal:2000wf,Alford:2007xm,Anglani:2013gfu} for reviews. At large and positive $\mu_I$ we expect  to populate the $u$ and $\bar d$ states with the color interaction inducing the formation of $\pi^+$ states that will eventually condense. A negative isospin chemical potential does  instead favor the formation of $\pi^-$ states. Whether these phases persist to the hadron gas surface of Figure~\ref{fig:phase_diagram} depends on the non-perturbative  properties of QCD.  This diagram is somehow our starting point to set the stage,  we~ will review it in Section~\ref{sec:conclusions} feeding in the up to date results. For the time being we note that there are two uncontroversial results: 
The critical temperature $T_c \sim 160$ MeV has been experimentally investigated  at RHIC and at LHC and precisely determined by LQCD simulations~\cite{wup_buda,hotqcd_2} to correspond to a analytic crossover.  The transition to the pion condensed phase at $T=0$ is a second order phase transition happening exactly at $\mu_I=m_\pi$~\cite{Son:2000xc,Kogut:2001id}. 


\subsection{Outside the Beta-Equilibrated  Sheet}

The three   chemical potentials, $\mu_B, \mu_I$ and $\mu_S$ are not three independent quantities because the weak interactions regulate the isospin and strangeness content of matter at a given baryonic density. Moreover, {matter must be electrically neutral and} the strong interactions can modify the dispersion laws of quasiparticles. Therefore, we should indicate in Figure~\ref{fig:phase_diagram} a  {\it beta-equilibrated  sheet} $\mu_I = f(\mu_B,T)$, where $f$ is some equation of state giving the  isospin asymmetry at any temperature and baryonic chemical potential. The beta-equilibrated  sheet corresponds to the configuration realized in long lived systems, as in compact stars. It is however instructive to consider configurations outside this surface, for three main reasons. The first is that we do not actually know $f$, except in  restricted  energy regions: at small energy scales by nuclear experiments, and at asymptotically large energy scales by perturbative QCD; the second is that  in the theoretical investigation we can turn off the weak interactions, thus~ we can compare the outcomes of  different theoretical methods outside the beta-equilibrated  sheet to test their robustness and consistency. Finally, working on the beta-equilibrated  sheet is interesting for studying the properties of dense nuclear matter in compact stars, but makes the problem so complicated that it is presently  hard to make  predictions.

In  the first papers discussing the  pion condensation~\cite{Migdal:1971cu,Migdal_1972}, Migdal tried to work on the beta-equilibrated  sheet  considering how nuclear matter mechanisms could make the  $\pi^0$ stable states. Then, different authors~\cite{Sawyer:1972cq, Scalapino:1972fu,1972PhLA...41..129K} considered the possible mechanisms for in-medium stabilization of charged pions by a  softening  of the pion spectrum by  the  $p$-wave  pion-nucleus interaction. The~ in-medium pion dispersion law  was assumed to be 
\be
\omega^2(k) = m_\pi^2 + 0.7 n m_\pi \omega + k^2(1- 6 n)\,,
\ee
where $n$ is the baryonic density in units of fm$^{-3}$. The exciting result is that a  gapless mode appears for $n>1/6 $ fm$^{-3}$, thus very close to the nuclear saturation density   $n_0 \sim 0.16$  fm$^{-3}$, at  a momentum $k = m_\pi/\sqrt{6 n -1}$. Therefore, a  transition to a superfluid phase was expected   just above the nuclear saturation density. However, it was pointed out by Migdal~\cite{Migdal:1973zm} that  the $\pi-\pi$ repulsive interaction may qualitatively change this result and he argued in favor of $\pi^0$ condensation as well as stable  $\pi^+ \pi^-$ molecules.  Still today we have not solved this problem, but most of the theoretical works are now directed to understanding what happens outside  the beta-equilibrated  sheet. This new approach has allowed Son and Stephanov~\cite{Son:2000xc}  to qualitatively and quantitatively assess the main properties of the pion condensed phase by the  use of a simple approach based on the modelization of QCD by chiral perturbation theory ($\chi$PT)~\cite{Son:2000xc,Kogut:2001id} .

Actually, there is a number of theoretical approaches that can be used.  In principle, any~ information on the  phase diagram of Figure~\ref{fig:phase_diagram} could be obtained introducing in the QCD action a chemical potential for the charge of interest. The QCD  Lagrangian turns to be
\be\label{eq:QCD_lagrangian}
\mathcal{L}_\text{QCD}=\bar{\psi}\left[\gamma^{\mu}\left(iD_\mu\right)  - M \right]\psi - \frac{1}4 F_{\mu\nu}^a F^{a,\mu\nu}\,,
\ee
where $F_{\mu\nu}^a$ with $a=1,\dots,8$ are the color gauge field strengths, $\psi^T=(u,d,s)$ is the spinor describing up, down  and strange quarks  (with suppressed color and spinorial indices), the bare quark masses are collected in the mass matrix
\be\label{eq:quark_mass_matrix}
M= \text{diag}(m,m,m_s)\,,
\ee
where we have assumed degenerate light quark masses,  and the covariant derivative 
\be\label{eq:covariant_quarks}
D_\mu  = \partial_\mu + i g A_\mu -\frac{i}{2} v_\mu\,,
\ee
includes both the minimal interaction with the gauge fields, $A_\mu$, and with the external source
\be\label{eq:vmu}
v_\mu=2 \mu \delta_{\mu 0}\,,
\ee
with the chemical potentials collected in the matrix
\be\label{eq:mu}
\mu=\text{diag}\left(\mu_u,\mu_d,\mu_s\right)= \frac{\mu_B-\mu_S}3 {\cal I} + \mu_I T_3 + \frac{2 \mu_S}{\sqrt{3}} T_8\,,
\ee
where $T_3$ and $T_8$  are the two diagonal $SU(3)$ generators,  $\cal{I}$  is the $3 \times 3$ identity matrix and  we have parameterized the quark  chemical  potentials as $\mu_{u,d} = \mu_B/3 \pm \mu_I/2$ and $\mu_s=\mu_B/3-\mu_S$. To determine the phase diagram in Figure~\ref{fig:phase_diagram} one should obtain the behavior of the chiral, pion and diquark  condensates, respectively given by
\begin{align}
\sigma & \propto \langle \bar \psi \psi \rangle\,,\label{eq:chiral_cond}\\
 \pi_a  & \propto \langle \bar \psi \sigma_a \gamma_5 \psi \rangle\,,\label{eq:pion_cond}\\
 \Delta  & \propto \langle  \psi C \gamma_5  \psi \rangle\,,\label{eq:diquark_cond}
\end{align}
where $\sigma_a$ with $a=1,2,3$ are the  Pauli matrices, as a function of $T$, $\mu_B$ and  $\mu_I$.

  

Given the nonperturbative character of QCD at the energy scales  of the various phase transitions in Figure~\ref{fig:phase_diagram}, the Lagrangian in Equation~\eqref{eq:QCD_lagrangian} is of little  direct use. Various theoretical approaches have been developed, including linear sigma models and chiral perturbation theory ($\chi$PT)~\cite{Baym:1978sz,Kaplan:1986yq, Dominguez:1993kr, Son:2000xc,Kogut:2001id,Birse:2001sn,Splittorff:2002xn,Loewe:2002tw,Loewe:2004mu,Loewe:2011tm,Mammarella:2015pxa,Carignano:2016rvs, Loewe:2016wsk,Carignano:2016lxe,Lepori:2019vec,Adhikari:2019mdk,Tawfik:2019tkp,Mishustin:2019otg}, the  
Nambu-Jona Lasinio (NJL) models~\cite{Barducci:1990sv,Toublan:2003tt, Barducci:2004tt,Barducci:2004nc, He:2005nk, Ebert:2005cs, Ebert:2005wr, Mukherjee:2006hq, He:2005sp, He:2006tn, Sun:2007fc, Andersen:2007qv,Abuki:2008tx, Abuki:2008wm, Mu:2010zz,Xia:2013caa,  Xia:2014bla, Chao:2018ejd, Khunjua:2018jmn,Khunjua:2019lbv, Khunjua:2019nnv,Avancini:2019ego,Lu:2019diy}, 
the quark-meson models~\cite{Klevansky:1992qe, Andersen:2014xxa, Adhikari:2016eef, Adhikari:2018cea,Andersen:2018osr,Andersen:2018qkq},  
the random~ matrix model~\cite{Klein:2003fy,Klein:2004hv}, the AdS/QCD model~\cite{Lv:2018wfq} and perturbative QCD (pQCD) (with diagrams resummation)~\cite{Graf:2015pyl, Andersen:2015eoa}.
A guiding role in this forest of theoretical approaches is played by the lattice QCD (LQCD) simulations~\cite{Alford:1998sd, Kogut:2002tm, Kogut:2002zg,Kogut:2004zg, Beane:2007es,Detmold:2008fn,Detmold:2008yn,Detmold:2011kw, Detmold:2012wc, Endrodi:2014lja, Janssen:2015lda, Brandt:2016zdy, Brandt:2018omg, Brandt:2017zck, Brandt:2018wkp}, which provide  a powerful  tool  for a numerical check and for exploring non-perturbative QCD. As we shall discuss in some detail in Section~\ref{sec:lattice}, the grand-canonical LQCD  simulations at finite baryonic density {and/or strangeness density} are hampered by the so-called sign problem, but are {feasible} at $\mu_B=\mu_S=0$ and  $\mu_I\neq 0$~\cite{Alford:1998sd}; moreover it is possible to simulate an  ensemble of kaons by the canonical LQCD approach~\cite{Detmold:2008yn,Detmold:2011kw}, corresponding to a system at nonvanishing strangeness density, {as discussed in more detail in Section~\ref{sec:lattice}}.   This~ places the study of the meson condensation  on a firmer ground with respect to the study of the phases at large baryonic density:   the various theoretical models (we will mostly focus on $\chi$PT and the NJL model) give  a qualitative and semiquantitative description of the meson condensed phase; the~ LQCD simulations  provide numerical evidence for the proposed  phase transitions and for the  meson  properties. The~ theoretical understanding  at large $\mu_B$  cannot count on  experimental data  nor on numerical  simulations. The~  phases realized at high baryonic densities could be relevant for dense stellar objects, but it is hard to obtain constraints on the microscopic properties of matter from the macroscopic properties of compact stars~\cite{Shapiro:1983du, Glendenning:1997wn}.


This review is organized as follows. In Section~\ref{sec:simple_modeling} we investigate the stability of pions in nuclear matter by a simplified non-interacting model description of hadronic matter. This  model, mainly used in the  1960s and in the 1970s,  serves as a guide for understanding by simple qualitative reasoning how meson condensation can occur and why it is unclear whether it be realized in compact stars or any other stellar object.  In Section~\ref{sec:group} we use an argument based on group theory to derive the phase diagram of the  meson condensed phases. This is a useful result because any other theoretical modeling is expected to reproduce this phase diagram. In Section~\ref{sec:modern} we report and compare the results obtained by three different  approaches: $\chi$PT, NJL and LQCD, showing that the obtained results are in qualitative and quantitative agreement for $T=0$ and $\mu_I \lesssim 2 m_\pi$. At nonvanishing temperature the three approaches give similar qualitative results, but more work is needed to  reconcile the  $\chi$PT and NJL methods  with the precise numerical results of the LQCD simulations.    We conclude, with a new discussion  of the QCD phase diagram, in Section~\ref{sec:conclusions}.

\section{The Early Works and Models}\label{sec:simple_modeling}

To understand why the pion condensation in dense hadronic matter is controversial~\cite{Migdal:1990vm} we  scrutinize the condition for the meson stabilization against weak decays by a simple non-interacting gas  approximation (NGA). This is  a mean field model  based on the assumption that neutrons, protons~ and electrons behave as independent Fermi  gases of quasiparticles. As we shall see below, the result of these kind of models is that fermions are favored with respect to mesons~\cite{1959ApJ...130..884C, 1960SvA.....4..187A, 1960AnPhy..11..393S, 1962SvA.....5..779A,Bahcall:1965zza}. The pion states can only be populated at non extreme densities   if the strong interaction modifies the pion spectrum.

\subsection{The Equilibrium Configuration}
The condensation mechanism of any type of particle relies on three basic requirements:
\begin{enumerate}
\item{ The particles must be bosons, as $^4$He atoms,  or boson-like,  as  Cooper pairs in the BCS theory}
\item{  The system has to be sufficiently cold: the particle condensation can be disrupted by the  thermal~ disorder}
\item{ The particles must be stable.   }
\end{enumerate}

Since mesons are bosons, they satisfy the first requirement. We can also imagine that they can be produced in a relatively cold environment, as in the core of neutron stars~\cite{Shapiro:1983du}. The third point is typically neglected in ultracold atom  physics, because experiments are  done with stable atoms~\cite{giorgini-review}. In~ contrast,  all mesons in vacuum are unstable. The point is whether medium effects can stabilize mesons or not.

The first papers discussing  stable pions in dense nuclear matter appeared in the 1960s~\cite{1959ApJ...130..884C, 1960SvA.....4..187A, 1960AnPhy..11..393S, 1962SvA.....5..779A}  considering a medium of catalyzed nuclear matter:   neutral matter consisting  of nucleons and leptons in electroweak equilibrium. 
In the Fermi gas approximation, the equilibrium distribution  is determined by the  Urca process 
\be\label{eq:Urca}
n \to p + e + \bar \nu_e \qquad
p + e \to n + \nu_e\,,
\ee
that with the assumption of neutrino transparency implies that 
\be\label{eq:chemicals}
\mu_n=\mu_p+\mu_e\,,
\ee
where $\mu_i$, with $i=n,p,e$, are the appropriate chemical potentials, see~\cite{Alford:2018lhf} for the temperature corrections to the Fermi gas approximation.
The electrical neutrality  requires that  
\be\label{eq:neutrality1}
n_p = n_e\,,
\ee
where $n_p$ and $n_e$ are the number densities of protons and electrons, respectively. 

In the NGA, the number density of each component is
\be\label{eq:ni}
n_i = 2 \int_0^{p_{f,i}}\frac{d^3 p}{(2 \pi)^3}= \frac{p_{f,i}^3}{3 \pi^2}\,, 
\ee
with $p_{f,i}$ the  Fermi momentum. The neutrality condition Equation~\eqref{eq:neutrality1} then implies that
the Fermi momenta of electrons and protons are the same, that is
\be\label{eq:pp-pe}
p_{f,p}=p_{f,e}\simeq \mu_e\,,
\ee
where in the last expression we have neglected the electron mass.  For free Fermi gases,  without~ in-medium effects, 
 we obtain using Equation~\eqref{eq:chemicals} that 
\be\label{eq:pn-pe}
p_{f,n} \simeq 2 \mu_e(\mu_e +\sqrt{\mu_e^2 + m_n^2})\,, 
\ee
where $m_n$ is the nucleon mass (we have neglected the small mass difference between protons and neutrons). Upon substituting these expressions in Equation~\eqref{eq:ni}   we can  express any number density as a function of the electron chemical potential; in particular,   the total number density as an increasing function of the electron chemical potential. We can now include  different hadronic states. We determine and compare, for reasons that will soon become  clear,  the threshold electron chemical potential for the appearance of $\pi^-$ and $\Sigma^-$. 

\begin{figure}[H]
\centering
\includegraphics[width=1.\textwidth]{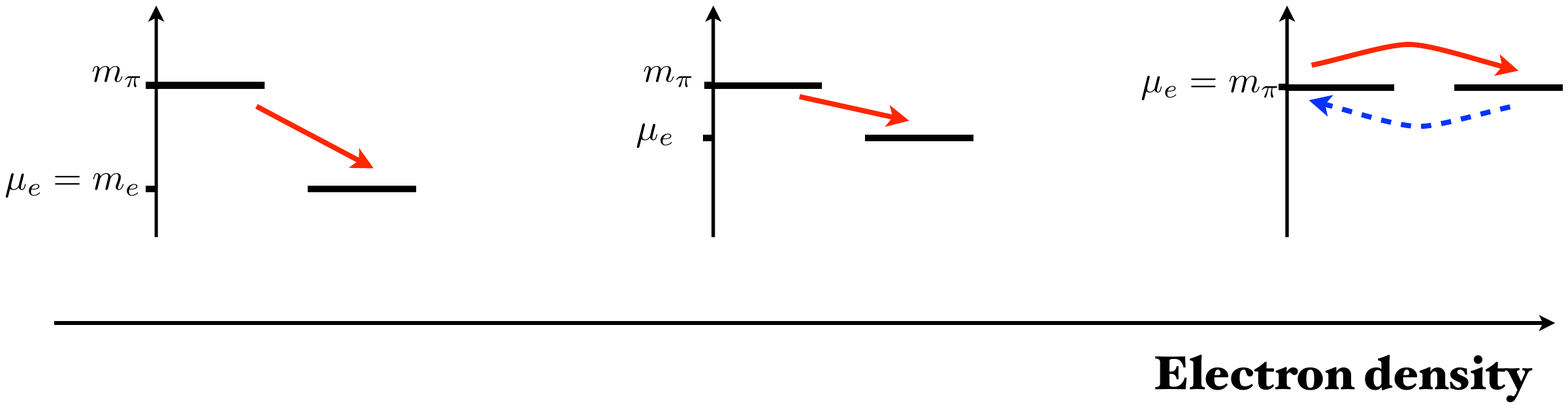}
\caption{{Sketch} of the Pauli blocking mechanism for the charged pion decay (solid red line)  induced by an increasing electron chemical potential.  With increasing electron density, the electron chemical potential grows.   The equilibrium configuration is reached for sufficiently high electron density, last~ figure on the right, with the pion decay process equilibrated by  the electron decay process (blue dashed line), corresponding to $\mu_e=m_\pi$.  }
\label{fig:pi_blocked}
 \end{figure}

In vacuum $\pi^-$ decays to leptons, mostly in muon and in its antineutrino. Since muons decay in electrons and neutrinos,  we  simplify the discussion considering the process  (solid red line in Figure~\ref{fig:pi_blocked})
\be\label{eq:pidecay}
\pi^- \to e^- + \bar\nu_e \,,
\ee
where we assume  that neutrinos are not trapped. This process can be Pauli blocked as depicted in the right side of  Figure~\ref{fig:pi_blocked}, meaning that it is in equilibrium with the electron decay process (dashed blue line in Figure~\ref{fig:pi_blocked})
\be\label{eq:e_decay}
e^- \to \pi^-  + \nu_e \,,
\ee
in the  configuration with
\be
\mu_{e \pi} = m_\pi \simeq 135~\text{MeV}\,.
\ee

We now proceed with a similar analysis for  the stability of the  $\Sigma^-$ baryonic resonance. In the quark model it is a $(dds)$ state with mass, $m_{\Sigma^-} \simeq 1.2$ GeV, weakly decaying  by \begin{equation}\Sigma^- \to n \pi^- \to n + e^- + \bar \nu_e\,, \ee in about $10^{-10}$ s. The $\Sigma^-$ states can be populated by the electron capture process  
\be\label{eq:sigma_production}
e^- + n  \to \Sigma^- + \nu_e \,,
\ee
with eventually an additional spectator neutron  to ensure the energy-momentum conservation. 
The~ equilibrium is reached at 
\be\label{eq:mu_e_sigma}
\mu_{e \Sigma}=\frac{2 m_{\Sigma^-} - \sqrt{m_{\Sigma^-}^2  + 3 m_p^2}}{3} \simeq 126 \text{ MeV}\,,
\ee
and since $\mu_{e \pi}>\mu_{e \Sigma}$  the $\Sigma^-$ states appear at a lower density  than the  $\pi^-$ states. This result is somehow surprising: at high density the heavier $\Sigma^-$ states are  favored over  the lighter mesonic $\pi^-$ states. This~ happens because the production channels of these two hadrons are different and because the $\Sigma^-$ is not much heavier than nucleons.  The density for the appearance of the $\Sigma^-$ states in the NGA is pretty large, about $4 n_0$. Once the  $\Sigma^-$ states are populated they must be included in the neutrality condition in Equation~\eqref{eq:neutrality1}, forbidding the appearance of pions up to $n\sim 300\, n_0$, see for instance~\cite{Bahcall:1965zza} and references therein.

\subsection{Including  in-Medium Effects}

Although in the free gas approximation the  $\pi^-$ is not energetically favored with respect to $\Sigma^-$,  the difference between the two critical chemical potentials is small,  $\mu_{e\pi}- \mu_{e\Sigma} \simeq 10 \text{ MeV} \ll m_\pi$. Thus,  a slight change of   their effective masses may invert this result:  charged pion states  could be populated~ first. 

To gain insight, we assume that the  the strong interactions cause a constant Fermi energy shift~\cite{Bahcall:1965zza}, independent of the particle momentum. In the NGA framework, for any baryon  $i$, we define the effective chemical potential
\be
\mu_{\text{eff},i} = \mu_i - \bar B_i\,,
\ee
where $\bar B_i$ is the constant chemical energy shift due to the strong interactions, while for leptons  $\bar B_\text{lepton}=0$.  The chemical equilibrium, Equation~\eqref{eq:chemicals}, now reads
\be\label{eq:effeq}
\mu_{\text{eff},n} = \mu_{\text{eff},p} + \mu_{e}\,, 
\ee
where we have assumed as before that neutrinos escape:  the medium does not trap neutrinos.  Turning to the problem of populating $\pi^-$  and $\Sigma^-$ states, the charged pion states are now populated (do not decay in leptons)  at
\be
\mu_{e \pi} = 
m_\pi - \bar B_\pi  \,,
\ee
thus a positive $B_\pi$ favors the appearance of $\pi^-$. 
The $\Sigma^-$ appears at the critical electron chemical potential   in Equation~\eqref{eq:mu_e_sigma} with
\be
m_{\Sigma^-} \rightarrow m_{\Sigma^-} - \bar B_{\Sigma^-}+\bar B_p\,,
\ee
therefore assuming small binding energies we find that pions appear first ($ \mu_{e \pi}  \leq \mu_{e \Sigma}$) for 
\be
\bar B_\pi - 0.47 ( \bar B_{\Sigma^-}-\bar B_p) \gtrsim 10 \text{ MeV} \,,
\ee 
which is a condition depending on three in-medium quantities.


The point of this simple model is that the appearance of pions may or may not happen depending on  quite small parameters  that  are not under quantitive control. In particular, it is not clear in which direction the medium effects go~\cite{Bahcall:1965zza}. Any result obtained in this way can hardly  stand firm against scrutiny and it is doomed to be troublesome,  because  based on a too naive  modeling and/or  on extrapolating  nuclear matter properties at least to  $2-3\, n_0$~\cite{Kaplan:1986yq}. 

\section{Group Theory Analysis}\label{sec:group}
In recent years a simpler approach to the mechanism of  meson condensation has been developed, disentangling it from  weak equilibrium and  strong interaction effects. We illustrate it for pions. Pions~ are an isospin triplet:  they are the three eigenstates of an $I=1$ multiplet with different $I_3$ projection. In vacuum the masses of the charged pions are degenerate, however this degeneracy is removed by the isospin chemical potential as in the Stark-Lo Surdo effect. In particular, we expect that the energy levels split as follows
\begin{align}
 E_{\pi^0} &= \sqrt{m_\pi^2+p^2}\,, \\
      E_{\pi^-}&=+ \mu_I + \sqrt{m_\pi^2+p^2}\,,\\
 E_{\pi^+}&=  - \mu_I+ \sqrt{m_\pi^2+p^2}\,,
\end{align}
{which are valid for $|\mu_I| <m_\pi$}, because  for $|\mu_I| >m_\pi$ one of the two charged mesons becomes massless. This is an indication of a spontaneous symmetry breaking (SSB), with the resulting  massless mode to be identified with the Nambu-Goldstone boson  (NGB)  associated to a broken generator. The best tool for exploring the (global) symmetry breaking is group theory. The good thing about this method is that it gives robust results. Its main  limitation is that it does not provide a microscopic mechanism for the occurrence of the symmetry breaking.

\subsection{Global Symmetries of QCD}

In  Figure~\ref{fig:breaking_pattern} we report the global symmetry breaking  pattern which is relevant for meson condensation.  
We start assuming  three flavor massless quarks  with $\mu_I=\mu_S=0$ described by an Hamiltonian, ${\cal H}$,  with global  symmetries
\be\label{eq:G_group}
\text{G}= \underbrace{SU(3)_L\times SU(3)_R}_{\displaystyle\supset U(1)_Q} \times U(1)_B 
\,,
\ee
where we have not included the $U(1)_A$ anomalous group and we have specified  that the electromagnetic  gauge group, $U(1)_Q$,  is a subgroup of the considered symmetries. The baryon symmetry group, $U(1)_B$,  can be broken by the formation of quark Cooper pairs in the CSC phase, see~ Figure~\ref{fig:phase_diagram}. In the following we will not consider this possibility, focusing on the chiral 
symmetry, $SU(3)_L\times SU(3)_R $, corresponding to the invariance of massless QCD with respect to    left- and right-handed rotations, {indicated respectively with  $U_L$ and $U_R$ in Figure~\ref{fig:breaking_pattern}},  of the quark fields.

\begin{figure}[H]
\centering
  \includegraphics[width=.8\textwidth]{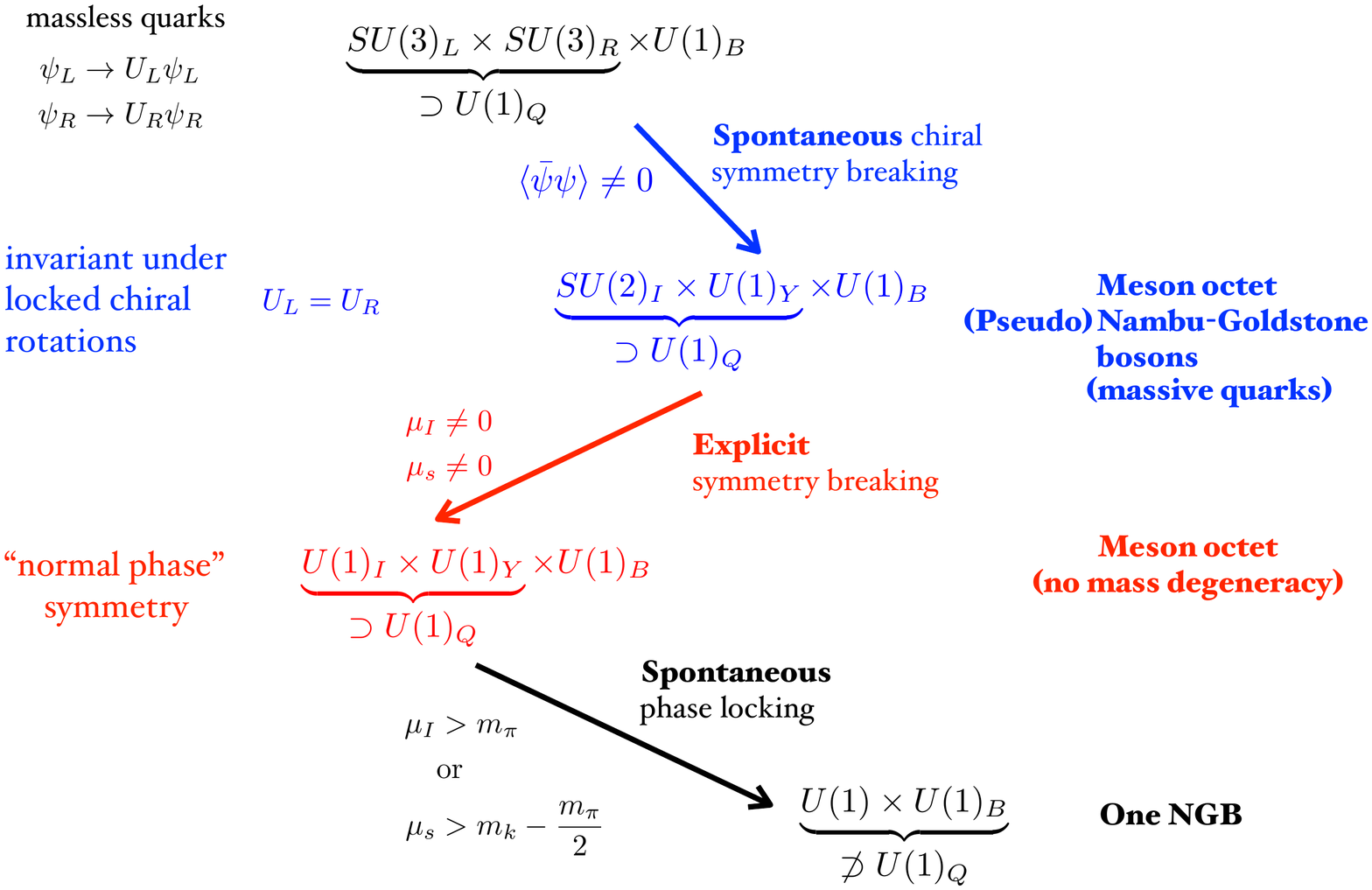}
\caption{Symmetry breaking path of three flavor quark matter; the arrows indicate the various  symmetry breakings. On the right we report the emergent low-energy degrees of freedom. The normal phase is defined by the corresponding symmetry group, see Equation~\eqref{eq:normal_phase}. See the text below for a through  description of the symmetry breakings.   }
\label{fig:breaking_pattern}
 \end{figure}

 In vacuum the chiral condensate locks the chiral rotation to the vectorial flavor group by the SSB
\be
\text{G} \to  \underbrace{SU(3)_V}_{\displaystyle\supset U(1)_Q}\!\!\!\times\, U(1)_B\,,
\ee
corresponding to the simultaneous rotations of left- and right-handed quark fields. The ground state is still invariant under rotation of quark flavors but the left and right handed rotations are now locked: {the  quark transformation leaving the vacuum invariant is $U_L=U_R$}. According to the Goldstone's theorem this symmetry breaking pattern  results in $8$ NGBs associated to the broken generators, see the left side of Figure~\ref{fig:splittings}.  Actually, the bare quark masses explicitly break the chiral symmetry, meaning that these modes are  massive pseudo NGBs, identified with the pseudoscalar meson octet. Assuming that the light quark masses are degenerate  the isospin symmetry is preserved, meaning that  the resulting symmetry is now (second row of Figure~\ref{fig:breaking_pattern})
 \be
 \underbrace{SU(2)_I \times U(1)_Y}_{\displaystyle\supset U(1)_Q}\times U(1)_B\,,
 \ee
where $SU(2)_I$ is the isospin symmetry group and  $U(1)_Y$  is the symmetry group generated by $T_8$. Thanks to the isospin invariance the pions are degenerate, with  mass $m_\pi \simeq 135$ MeV.  The kaons are grouped in two isodoublets with degenerate masses $m_K \simeq 500$ MeV.
The kaon masses differ from the  pion masses because kaons involve a strange quark.  These meson masses are reported on the central part of Figure~\ref{fig:splittings}. The last member of the octet, which is not reported in Figure~\ref{fig:splittings},  is  the $\eta$ field,  with~ mass  
\be
    m_{\eta} \simeq    \sqrt{\frac{4 m_K^2-m_\pi^2}{3}}\,,
\ee
by the Gell Mann-Okubo relation. 

\begin{figure}[H]
\centering
\includegraphics[width=10.cm]{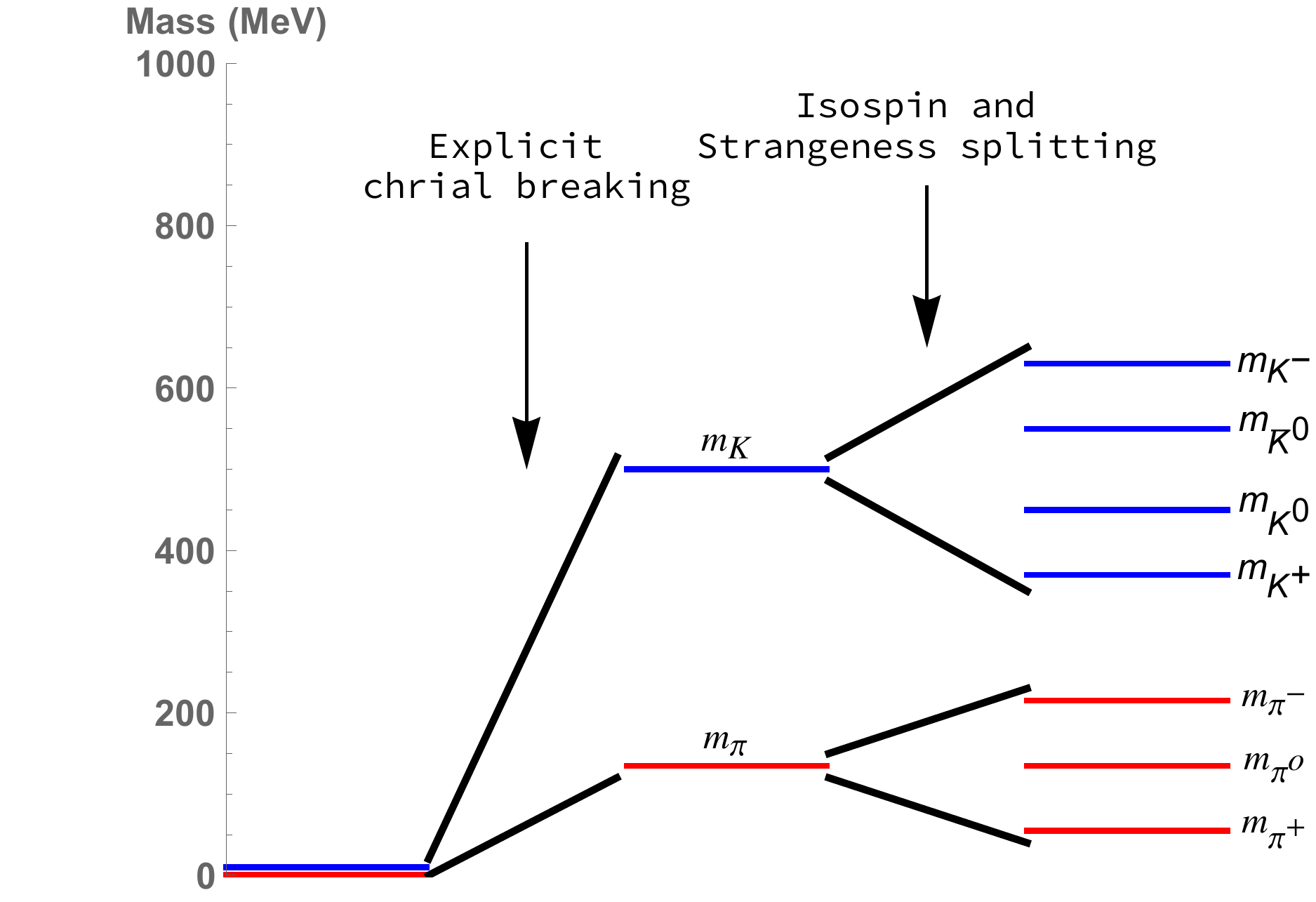}
\caption{Sketch of the energy levels of pions and kaons. On the left we assume massless quarks and vanishing chemical potentials.  The effect of the explicit symmetry breaking induced by the quark masses is shown in the central region. On the right we show the mass splitting within isomultiplets,  obtained~  for $\mu_I=80$ MeV and $\mu_S=90$ MeV. The   isospin and  the strangeness chemical potentials completely remove the level degeneracy. }
\label{fig:splittings}
\end{figure}

The chemical potentials  induce  two different symmetry breakings, one is explicit, at the Lagrangian level, while the second one is a SSB, causing the  meson condensation.  At the Lagrangian level, the chemical  potentials  explicitly break the Lorentz boost invariance, due to the  presence of a privileged reference frame,  the one in which particles are at rest with the medium.
Space rotations and translations are instead unaffected: the considered medium is isotropic and homogeneous.   The~ chemical potentials also explicitly break the charge symmetry, meaning that charged conjugated states will have different masses. This effect can also be seen in  a more detailed way scrutinizing how the  chemical potentials in Equation~\eqref{eq:mu} break  the flavor symmetry. Since 
\be
[{\cal H},T_3] =0 \qquad \text{and}\qquad  [{\cal H},T_8] =0\,,
\ee
while the commutator of the Hamiltonian with any other $SU(3)$ generator is nonzero, 
the explicit symmetry breaking is  (third row in Figure~\ref{fig:breaking_pattern})
 \be\label{eq:normal_phase}
\underbrace{SU(2)_I \times U(1)_Y}_{\displaystyle\supset U(1)_Q}\times U(1)_B \to \text{N}= \underbrace{U(1)_I\times U(1)_Y }_{\displaystyle\supset U(1)_Q}\times U(1) _B
 \ee
 where $U(1)_I$ and $U(1)_Y$ are the symmetries generated by $T_3$ and $T_8$, respectively. The remaining  symmetry group  can be viewed as generated by the independent phase rotations of the three flavor fields. This residual symmetry is extremely important, because it proves  possible a SSB and a phase transition between 
 the {\it normal phase}, characterized by the symmetry N in~Equation \eqref{eq:normal_phase},
 to a superfluid phase, with a reduced global symmetry.  The  meson condensation, last part of the diagram in Figure~\ref{fig:breaking_pattern}, is indeed related to the SSB  that locks the phases  of quarks with different flavors. It is  induced by a large $\mu_I$ or $\mu_S$.

\subsection{Phases of Condensed Mesons}
We can figure out the conditions for the final SSB of  Figure~\ref{fig:breaking_pattern} by inspecting the meson mass spectrum. The isospin and strangeness chemical potentials remove the degeneracy within  isomultiplets and between states with different hypercharge. Since 
\be
[{\cal H},T_3] =0 \qquad \text{and}\qquad  [{\cal H},T^2] =0\,,
\ee
in the normal phase we can still group eigenstates as isospin multiplets. By the lemmas of Schur, these states will not be degenerate in mass: there is   a mass splitting within members of isomultiplets, proportional to $\mu_I$, and there is a mass splitting between states with different hypercharge,  proportional to $\mu_S$. The only unaffected states  are the $\pi_0$ and $\eta$ states, because they have both vanishing isospin  and hypercharge. Since the mass splitting is given by the corresponding charges we have that
\begin{align}
    m_{\pi^0}&=m_\pi\,, \\
    m_{\pi^\pm}&=m_\pi \mp \mu_I\,,\label{eq:pi+-masses}\\
    m_{\eta}&=    \sqrt{\frac{4 m_K^2-m_\pi^2}{3}}\,, \\
     m_{K^\pm}&=m_K\mp\frac12 \mu_I\mp\mu_S\,,  \\
    m_{K^0/\bar K^0}&=m_K\pm\frac12 \mu_I\mp\mu_S \,,
\label{eq:norm_masses}
\end{align} 
where  $m_\pi$ and $m_K$  indicate the meson masses for $\mu_I=\mu_S=0$.

To elucidate the mass splittings  we show on the right of Figure~\ref{fig:splittings} the results obtained  for $\mu_I=80$~ MeV and $\mu_S=90$ MeV. As discussed above, the first mass splitting is induced by the nonvanishing quark masses. The second  splitting is due to the considered values of the chemical potentials, which~ completely remove the   level degeneracies. The shown  hierarchy is only one of the possible ones: different values of $\mu_I$ and/or $\mu_S$ may  imply different   level splittings, possibly with   kaons  lighter than  pions.

The pion masses are independent of $\mu_S$ and we see from Equation~\eqref{eq:pi+-masses}  that for 
\be
\mu_I = |m_\pi| \,,
\ee
 one of the charged  mesons becomes massless. Analogously,  for $|\mu_S|=m_K-|\mu_I|/2$ one of the kaons becomes massless. The normal phase will persists until no mode is massless, that is for \be
|\mu_I| < m_\pi \qquad \text{and} \qquad |\mu_S| < m_k -\frac{|\mu_I|}2\,,
\ee
which corresponds to the irregular hexagon (solid red line) in Figure~\ref{fig:phases}. 
\begin{figure}[H]
\centering
\includegraphics[width=8.cm]{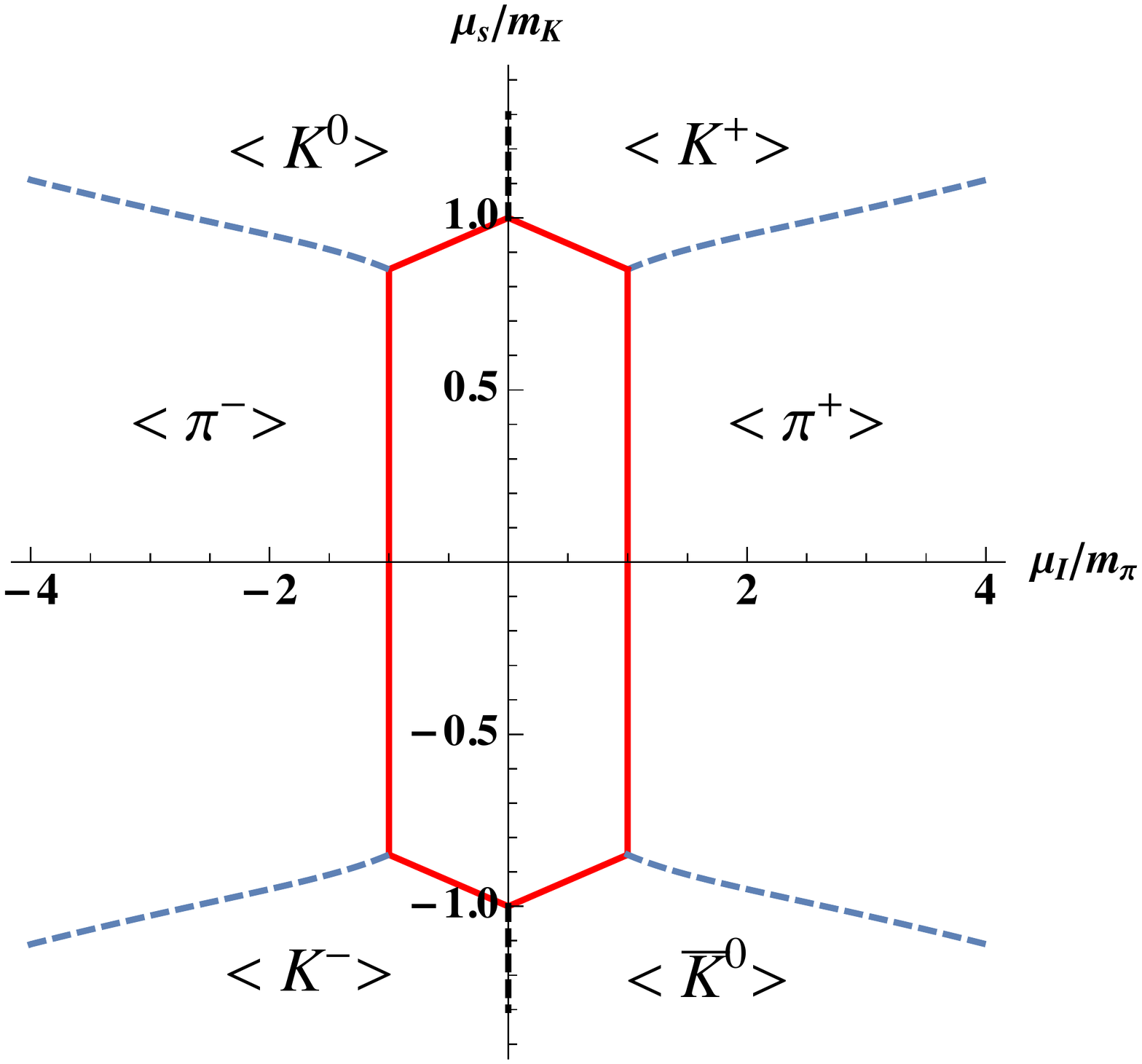}
\caption{Phase diagram of the meson condensed phase, see~\cite{Kogut:2001id}. The irregular hexagon (solid red line), obtained from the meson mass spectrum, corresponds to the second order phase transitions between the normal phase and the meson condensed phases. The dashed lines have been drawn by hand and represent the expected first order phase transitions between two different meson condensed phases.}
\label{fig:phases}
\end{figure}
This phase diagram was first derived in~\cite{Kogut:2001id} in the $\chi$PT framework. At the boundary of the hexagon we expect a massless mode, signaling a SSB.  This massless mode  eventually condenses if the temperature is below the relevant critical temperature. When meson condensation occurs, the~ vacuum will have a nonvanishing charge associated to  one of the non-diagonal generators of $SU(3)$ characterizing the condensed meson.

For the description of the broken phases, we first note that the $\mu_I$ and $\mu_S$ Lagrangian terms are proportional to $T_3$ and $T_8$, respectively, see Equation~\eqref{eq:mu}. Since these two generators form the Cartan subalgebra of $SU(3)$,  it turns out  useful to consider the three ${\cal SU}(2)$ Lie subalgebras of ${\cal SU}(3)$ generated~ by
\begin{align}
T_\pm\,, T_3 \,, \qquad   
V_\pm\,, V_3  \,,\qquad   
U_\pm\,, U_3 \,, 
\end{align}
where the step operators are respectively

\begin{align}
T_\pm = T_1 \pm i T_2  \,, \qquad
V_\pm = T_4 \pm i T_5 \,,  \qquad
U_\pm = T_6 \pm i T_7 \,,  
\end{align}
and 
 \begin{align}
Y  = \frac{2}{\sqrt{3}}T_8\,, \qquad
K= T_3 - \frac{1}{\sqrt{3} }T_8\,,\qquad
Q  =   T_3 + \frac{1}{\sqrt{3} }T_8\,,\label{eq:charges}
\end{align}
are the corresponding  weights.

Let us now consider the case in which one of the charged pions, say the $\pi^+$,    becomes massless and condenses.  The  spontaneously broken generator is $T_+$, which locks the phases of up and down quarks. Since $[T_+,Y]=0$,
it follows the SSB
\be
 \underbrace{U(1)_I\times U(1)_Y}_{\displaystyle\supset U(1)_Q} \times U(1) _B \to \underbrace{U(1)_Y\times U(1)_B}_{\displaystyle\not\supset U(1)_Q}\,,
\ee
meaning that the massless mode is the NGB associated to $U(1)_I$ breaking. We remark that  this  superfluid mode  actually signals the transition to a superconducting phase, because the $U(1)_Q$ symmetry is broken. 
Since  $[T_+,Q] \neq 0$ the charged pion states can mix  and indeed the
massless mode coincides with the $\pi_+$ only at the phase transition point. In general it will be a superposition of the two charged pion states. Analogous results hold if the  $\pi^-$ becomes massless and condenses, with~ $T_-$ the broken generator.  

The second possibility is that the  $K^+$ becomes massless and condenses;  the spontaneously broken generator is $V_+$, which locks the phases of up and strange quarks.  In this case $[V_+,K]=0$,
with the   resulting  SSB
\be
\underbrace{U(1)_I\times U(1)_Y }_{\displaystyle\supset U(1)_Q}\times U(1) _B \to  \underbrace{U(1)_{K}\times U(1)_B}_{\displaystyle\not\supset U(1)_Q}\,,
\ee
meaning that there is a residual global symmetry associated to the weight  $K$. In this case  the isospin and hypercharge phases  are locked. The system is a kaon superconductor, because the condensed meson is charged. Analogous results hold if the  $K^-$ becomes massless and condenses.

The third case is the condensation of one of the neutral kaons. If the $K^0$ becomes massless and condenses, then the spontaneously broken generator is $U_+$, with  down-strange quark phase locking. Since $[U_+,Q]= 0$,
the resulting  SSB is 
\be
\underbrace{U(1)_I\times U(1)_Y }_{\displaystyle\supset U(1)_Q}\times U(1) _B \to U(1)_Q \times U(1)_B\,,
\ee
meaning that the electromagnetic gauge field is unbroken. Therefore, the system is a kaon superfluid where the superfluid mode is given by the mixing of the two neutral kaons.

Let us analyze the order of the  transition lines in Figure~\ref{fig:phases}.  The $\chi$SB and the  meson condensation mechanisms are independent:  the $\chi$SB is related to the locking of the phases of left- and right-handed quarks, while the meson condensation locks the phases of quarks with different flavors. This means that the chiral condensate and the meson condensate can coexist, therefore  the irregular hexagon should be a second order phase transition line. On the other hand, it is not possible to have the simultaneous condensation of say $\pi^+$ and $K^+$ because both condensates involve an up quark. In group theory this is related to the fact that   the two generators $T_+$ and $V_+$ do not commute. We expect  that  the presence of a meson  condensate excludes the other, implying the first order phase transition shown in Equation~\eqref{fig:phases} as dashed lines. 
The simultaneous condensation can only happen at the  vertices of the hexagon 
or  close to the first order phase transition lines if   inhomogeneous phases are realized. This~ has been preliminarily explored by canonical LQCD simulation in~\cite{Detmold:2011kw} and by $\chi$PT   in~\cite{Lepori:2019vec}.


To summarize,  any meson condensate tilts the  vacuum in a certain direction  having a residual $ U(1)_B \times U(1) $ symmetry which is generated by the baryonic charge and by one of the weight operators in Equation~\eqref{eq:charges}. The low-energy spectrum consists of one NGB. This NGB  has the same quantum numbers of one of the standard mesons at the phase transition point, but then in the superfluid phase it   mixes with  its charged conjugated state as shown in Figure~\ref{fig:mixing}, see also~\cite{Mammarella:2015pxa}.

\begin{figure}[H]
\centering
\includegraphics[width=8.cm]{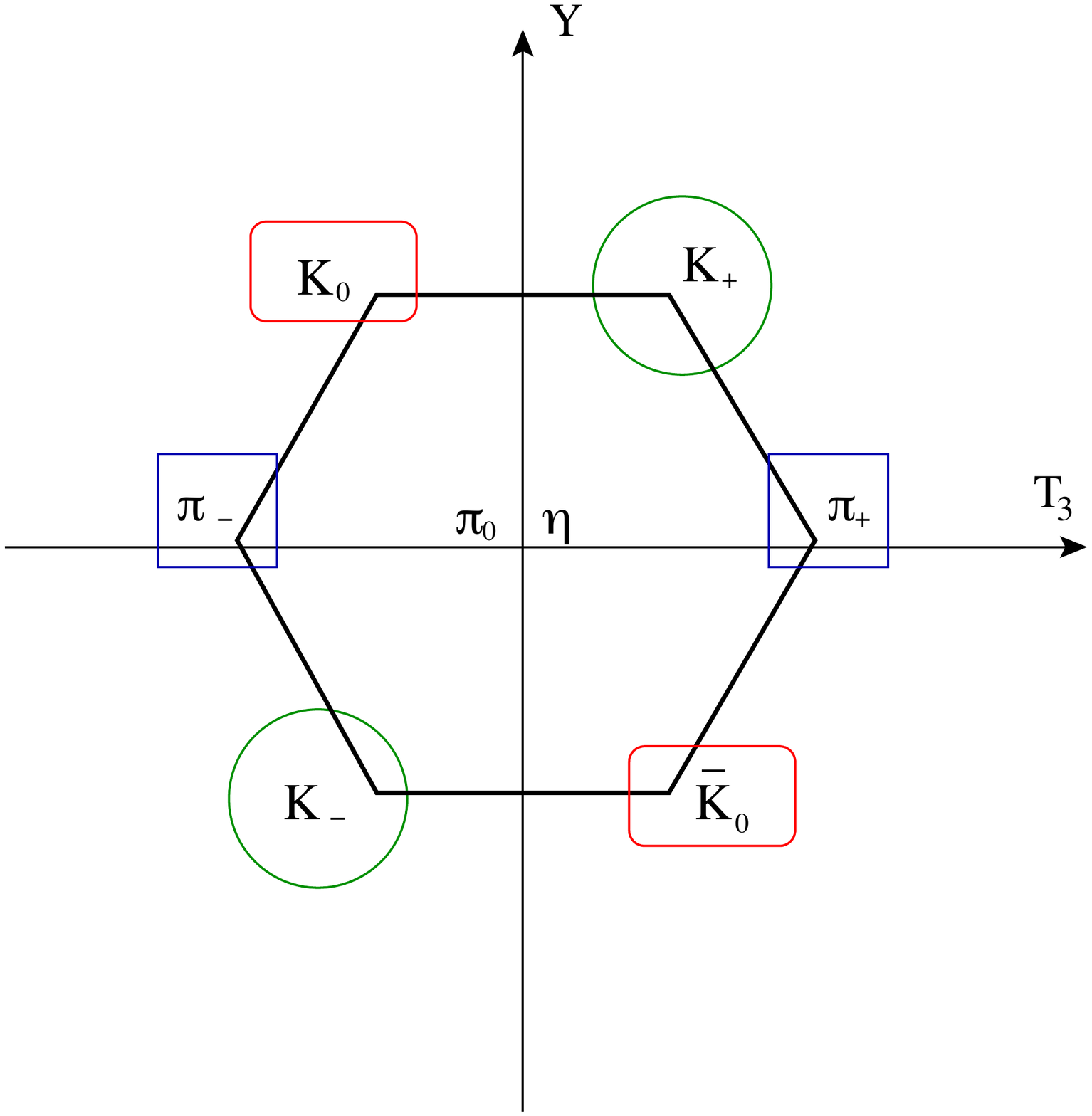}
\caption{Weight diagram of the mesonic octet. We have enclosed with the same symbol   the  states  that  can mix in the meson condensed phase. They have the same  $T$-spin, $U$-spin and $V$-spin quantum numbers.   The $\pi_0$ and the $\eta$  are not simultaneous   $T$-spin, $U$-spin and $V$-spin eigenstates; their mixing depends on the spontaneously induced charge of the vacuum, see~\cite{Mammarella:2015pxa} for more details.  }
\label{fig:mixing}
\end{figure}



\section{Modern Approaches}\label{sec:modern}

We now present two different approaches to the meson condensation, one is based on the effective field theory $\chi$PT description of  mesons, the second on the NJL modelization of the strong interaction by a contact term. Then we compare the results of  these methods with those of the pertinent LQCD numerical simulations. We restrict to  consider an homogeneous and static medium; for inhomogeneous phases see~\cite{Sadzikowski:2002iy, Anglani:2013gfu,Buballa:2014tba, Carignano:2017meb, Andersen:2018osr}. 

\subsection{Chiral Perturbation Theory}\label{sec:chiral}

The $\chi$PT Lagrangian is an extremely powerful tool  for systematically describe the strong interactions between hadrons~\cite{Weinberg:1978kz, Gasser:1983yg, Georgi:1985kw, Leutwyler:1993iq, Ecker:1994gg, Leutwyler:1996er,  Pich:1998xt, Scherer:2002tk, Scherer:2005ri}. Remarkably, $\chi$PT can also be used to study a variety of gauge theories with isospin asymmetry, including 2 color QCD with different flavors~\cite{Kogut:1999iv,Kogut:2000ek, Hands:2000ei,Kogut:2001na, Brauner:2006dv,Braguta:2016cpw, Adhikari:2018kzh}. Quite~ generally, any $\chi$PT realization  is  based on two key ingredients:  the global symmetries of the studied theory and an appropriate  low momentum expansion.

The relevant  global symmetry of QCD for constructing the $\chi$PT Lagrangian is the chiral symmetry
\begin{equation}SU(N_f)_L\times SU(N_f)_R\,,\ee  with $N_f$ the number of relevant quark flavors. 
The meson fields are collected in the unimodular $\Sigma$ field, transforming  as
\be
\Sigma \to  L \Sigma R^\dagger\,, 
\ee
where $L \in SU(N_f)_L$ and $R \in SU(N_f)_R$.  Based on the chiral  symmetry and this transformation property one builds the most general Lagrangian  at the given order in the momentum expansion, assuming
that the meson momenta  satisfy
 \be
p \ll \Lambda_\chi \,,
 \ee
where $ \Lambda_\chi \sim 1$ GeV is the $\chi$PT breaking scale.  At each order in the momentum expansion the chiral symmetry  fixes the form of the various Lagrangian pieces but  the pre-factors, the so-called low energy constants (LECs),  must be determined by different means. The leading   $ {\cal O}(p^2)$  $\chi$PT   Lagrangian~\cite{Kogut:2001id,Gasser:1983yg,Scherer:2002tk}  describing the  in-medium pseudoscalar mesons is given by  
\be\label{eq:chiral_Lagrangian}
{\cal L} = \frac{f_\pi^2}{4} \text{Tr} ({\cal D}_\nu \Sigma\, {\cal D}^\nu \Sigma^\dagger) + \frac{f_\pi^2}{4} \text{Tr} (X \Sigma^\dagger + \Sigma X^\dagger )\,,
\ee
where  the trace is in flavor space and the auxiliary field $X$ transforms as $\Sigma$. The locking of the chiral~ rotations to the vector  $SU(N_f)_V$ group is induced by the vev of $X$, see for example the discussion in~\cite{Georgi:1985kw,Scherer:2002tk}, which is usually written as $\langle X \rangle =2 B M$ with $M$ the quark mass matrix defined in Equation~\eqref{eq:quark_mass_matrix}.  The two LECS,  $B$ and  $f_\pi$,  can be fixed by  the vacuum properties~\cite{Gasser:1983yg, Leutwyler:1993iq, Ecker:1994gg,  Scherer:2002tk, Scherer:2005ri}; for instance $B$ from the mass relations   $m_\pi^2=2 B m $ and 
$m_K^2=B (m+m_s)$, while $f_\pi \simeq 93$ MeV from the  weak  pion decays.

The adjoint covariant derivative in Equation~\eqref{eq:chiral_Lagrangian} takes into account the minimal coupling of~ the~ meson fields with  gauge fields,   external currents and the effect of different chemical potentials~\cite{Gasser:1983yg, Leutwyler:1993iq,Leutwyler:1996er}. Pretty much as the covariant derivative in the quark sector, see Equation~\eqref{eq:covariant_quarks}, we~ define it as
\be\label{eq:covariant_mesons}
{\cal D}_\nu \Sigma = \partial_\nu\Sigma + \frac{i}{2} [v_\nu,\Sigma] \,,
\ee
where the external current, $v_\nu$,  is given in Equation~\eqref{eq:vmu}; here we can  take  $\mu=\text{diag}(\mu_I/2,-\mu_I/2,-\mu_S)$ because mesons have no baryonic charge. This  leading order (LO) $\chi$PT Lagrangian  is sufficient to accurately describe the phase structure of QCD at   $\mu_I \lesssim  2 m_\pi$~\cite{Son:2000xc,Kogut:2001id,Carignano:2016lxe},  including   finite temperature effects~\cite{Splittorff:2002xn,Loewe:2002tw, Loewe:2004mu} and pion in-medium stability~\cite{Mammarella:2015pxa,Loewe:2016wsk}.

\subsubsection{Ground State}
The ground state can be variationally determined by replacing   $\Sigma \to \bar \Sigma$ in Equation~\eqref{eq:chiral_Lagrangian},
where $\bar \Sigma$ is the time independent and homogeneous vacuum expectation value of the meson fields.
The resulting  static Lagrangian
\be\label{eq:chiral_Lagrangian_static}
\mathcal{L}_\text{static}=-\frac{f_\pi^2}{4} \Tr \left( \[\mu, \bar{\Sigma}\phantom{^\dag}\hspace{-.1cm} \] \[\mu,\bar{\Sigma}^\dag \] \right)+\frac{f_\pi^2 B}{2} \Tr\left(M (\bar{\Sigma}+\bar{\Sigma}^\dag)\right)\,, 
\ee
must be maximized to determine the vacuum configuration.
Specifically, one can parameterize the most general $SU(3)$  vev as
\be\label{eq:general_vev} 
\bar{\Sigma} = e^{i \alpha \bm n \cdot \bm \lambda}\,,
\ee
where $\lambda_i$, for $i=1,\dots,8$ are the Gell-Mann matrices,  and the tilting angle $\alpha$  and the eight-dimensional unit vector $\bm n$ are the variational parameters that should    be obtained maximizing Equation~\eqref{eq:chiral_Lagrangian_static}. The~ normal phase is easily described by  $\alpha=0$, thus $ \bar \Sigma =  \bar \Sigma_N = I$, but in general one should maximize Equation~\eqref{eq:chiral_Lagrangian_static} with respect to eight independent parameters, which is a rather formidable task. As far as I know, nobody has ever considered this procedure, and for a good reason.  From~ the insight gained in Section~\ref{sec:group} by the group theory analysis, we expect that in each different meson condensed phase the vacuum is rotated in a specific way. In the pion condensed phase the only nonvanishing components should be  $(n_1, n_2)$, in the charged kaon condensed phase  $(n_4, n_5)$, while~ in the neutral kaon condensed phase  $(n_6, n_7)$. One can further simplify the ansatz observing that the potential must have a flat direction,   the one  spanned by the NGB. 

To elucidate these aspects let us first focus on the $N_f=2$ case, with the most general ansatz 
\be\label{eq:pion_condensate}
\bar \Sigma = e^{i \alpha \bm n \cdot \bm \sigma}= \cos \alpha + i {\bm n} \cdot {\bm \sigma} \sin\alpha\,,
\ee 
where  $\bm n = (n_1,n_2,n_3)$. The static Lagrangian now reads
\be\label{eq:L_static_2}
{\cal L}_\text{static}= f_\pi^2 m_\pi^2 \cos\alpha +  \frac{f_\pi^2}2 (\sin\alpha)^2 \mu_I^2 (1-n_3^2)\,,
\ee
showing that it is independent of $n_1$ and $n_2$, corresponding to the flat direction spanned by the NGB   that interpolates between the $\pi_+$ and $\pi_-$ fields, see Section~\ref{sec:group}. For maximizing the Lagrangian one has to take $n_3=0$, then we have the freedom to take say $n_1=0$ and $n_2=1$. In summary,  the ground state has vanishing  projection along the isospin direction while the  rotations around the direction of the chemical potential leave the vacuum invariant.

We now return to  the three-flavor case. In the three different meson condensed  phases  we expect that it suffices to take  only one entry of the unit vector $\bm n$  nonzero. In particular,   in the pion condensed~ phase 
\begin{align} \bar{\Sigma}_{\pi}&= e^{i \alpha \lambda_2}  = \left( \begin{array}{ccc}
                            \cos \alpha & \sin \alpha & 0 \\
                            -\sin \alpha & \cos \alpha & 0\\
                            0 & 0 & 1
                            \end{array} \right) = \frac{1+2 \cos\alpha}{3} I + i \lambda_2 \sin\alpha+ \frac{\cos\alpha-1}{\sqrt{3}}\lambda_8\,,
                            \label{eq:pion_condensate}
\end{align}
while  in the charged kaon condensed phase 
\begin{align}  \bar{\Sigma}_K
=&e^{i \alpha \lambda_5}=\left( \begin{array}{ccc}
                            \cos \alpha & 0 & \sin \alpha  \\
                            0 & 1 & 0\\
                            -\sin \alpha & 0 & \cos \alpha
                            \end{array} \right)=
\frac{1+2 \cos\alpha}{3} I+ \frac{\cos\alpha-1}{2\sqrt{3}}\left(\sqrt{3}\lambda_3-\lambda_8\right) + i \lambda_5 \sin\alpha\,,\label{eq:kaon_condensate}
\end{align}
finally in the neutral kaon condensed phase
\begin{align}  \bar{\Sigma}_{K_0}
=&e^{i \alpha \lambda_7}=\left( \begin{array}{ccc}
                            1 & 0 & 0 \\   
                            0 & \cos \alpha & \sin \alpha\\
                            0 & -\sin \alpha & \cos \alpha
                            \end{array} \right)= \frac{1+2 \cos\alpha}{3} I+ \frac{1-\cos\alpha}{2\sqrt{3}}\left(\sqrt{3}\lambda_3+\lambda_8\right) + i \lambda_7 \sin\alpha\,,\label{eq:neutral_kaon_condensate}
\end{align}
where $\alpha$  assumes different values in the three different  phases. 

To allow the transition between the three phases one can consider the more general ansatz
\be\label{eq:general_ansatz}
\bar \Sigma = e^{-i \gamma \lambda_2} e^{-i \beta \lambda_7}  e^{i \alpha \lambda_2} e^{i \beta \lambda_7}e^{ i \gamma \lambda_2}\,,
\ee
with $\alpha,\beta,\gamma \in [0,\pi/2]$ are three different angles. Once again, the normal phase corresponds to $\alpha=0$ and it is insensitive to the values of $\beta$ and $\gamma$. The pion condensed phase corresponds to $\beta=0, \gamma=0$, the $K^+$ condensed phase to $\beta=\pi/2, \gamma=0$,  the $K^0$ condensed phase to $\beta=\pi/2, \gamma=\pi/2$. Any value of $\beta$ and $\gamma$ different from the above ones   indicates a phase with simultaneous  condensation of two or more meson fields. We have argued in Section~\ref{sec:group} that this is not the case. This has been explicitly shown in $\chi$PT for the $\pi^+$ and $K^+$ condensation phases~\cite{Kogut:2001id}. In this case we can simplify the ground state ansatz by taking $\gamma=0$ in Equation~\eqref{eq:general_ansatz}, obtaining the same ansatz of~\cite{Kogut:2001id}. Any value 
$\beta \in (0,\pi/2)$ corresponds to the simultaneous $\pi^+$-$K^+$ condensation. The analysis of~\cite{Kogut:2001id} shows that at the phase transition   the angle $\beta$ discontinuously jumps from $0$ to $\pi/2$ resulting in a first order phase transition (as depicted in Figure~\ref{fig:phases}): no simultaneous condensation is possible. 

Below we summarize the main properties of the possible phases obtained by   the $\chi$PT  analysis. The phase diagram is  in agreement with  Figure~\ref{fig:phases}, indeed it was first derived by $\chi$PT in~\cite{Kogut:2001id}. Given the symmetry of the phase diagram we focus on the  $\mu_I >0$ and $\mu_S >0$ part of  Figure~\ref{fig:phases} characterized by the $\pi^+$ and $K^+$ condensates. The red solid line is  exactly determined, while the $\chi$PT prediction  for the dashed blue line is 
\begin{equation}
 \mu_S =
\frac{-m_\pi^2+\sqrt{(m_\pi^2-\mu_I^2)^2+4 m_K^2 \mu_I^2} }{2 \mu_I}\,.
\ee

We also report below the relevant thermodynamic quantities.  Once $\bar \Sigma$ is determined, the pressure is obtained from  
\be
P = -{\cal L}_\text{static}(\bar \Sigma)\,,
\ee
and then the number densities and the energy density follow from the thermodynamic relations
\begin{align}
n_{I,S} = \frac{\partial P}{\partial \mu_{I,S}}\,,\qquad 
\epsilon=  \mu_I n_I + \mu_S n_S - P\,.
\end{align}

Finally one can obtain the equation of state (EoS) by appropriately expressing the chemical potentials as a function of the pressure.

\begin{itemize}
\item The normal phase is  favored for 
\begin{align} \mu_I&<m_\pi \,,\nonumber \\ 
\mu_S&<m_K-\frac12 \mu_I\,,
\end{align}
with the trivial vev
$\bar{\Sigma}_N=\text{diag}(1,1,1) $. 
The nonvanishing condensates are the three  chiral condensates, see Equation~\eqref{eq:chiral_cond},
\be
\sigma_u = \sigma_d = \sigma_s \equiv \sigma_0 \,,
\ee
where the subscript indicates the quark flavor and  $\sigma_0$ is the value of the chiral condensate in vacuum. In the normal phase the values of the chiral condensates are not affected by the chemical potentials. The pressure 
is given by
\begin{align}
P =\left\{ \begin{array}{ll} f_\pi^2 m_\pi^2 & \text{for  }  N_f=2 \,,\\
f_\pi^2 m_\pi^2 \left(\frac{1}2 + \frac{m_K^2}{m_\pi^2}\right) & \text{for  }  N_f=3 \,,\end{array}\right. 
\end{align}
and thus the isospin and strangeness number densities vanish, that is
\be
n_I = n_S = 0\,.
\ee

\item The $\pi^+$ condensed phase is  favored for 
\begin{align} \mu_I&>m_\pi\,, \nonumber \\
 \mu_S&<
\frac{-m_\pi^2+\sqrt{(m_\pi^2-\mu_I^2)^2+4 m_K^2 \mu_I^2} }{2 \mu_I}\,,
\end{align}
resulting in the vacuum in Equation~\eqref{eq:pion_condensate}
with 
\be\label{eq:alpha_pi}
    \cos \alpha_{\pi}=\left(\frac{m_\pi}{\mu_I}\right)^2\,,
    \ee  
determined by maximizing the static Lagrangian in Equation~\eqref{eq:L_static_2}. The condensates are given by
    \begin{align}
\sigma_u &= \sigma_d = \sigma_0 \cos  \alpha_{\pi} \qquad
\sigma_s = \sigma_0  \nonumber \\
\langle \pi^+ \rangle & =  \sigma_0 \sin  \alpha_{\pi} \qquad 
\langle K^+ \rangle  =  0\,,
 \end{align}
and    the  pressure produced by the condensation of pions is given by~\cite{Son:2000xc,Kogut:2001id} \be
P= \frac{f_\pi^2\mu_I^2}{2} \left(1- \frac{m_\pi^2}{\mu_I^2}\right)^2 \,, \label{eq:pressure-pions}
\ee
where the normal phase pressure has been subtracted.   This expression is valid for both $N_f=2$ and  $N_f=3$; it is of course insensitive to the  kaon mass and  the strange quark chemical potential.
The number densities are
\begin{align}
n_{I} = f_\pi^2 \mu_I \left(1-\frac{m_\pi^4}{\mu_I^4}\right)\qquad 
n_s  =0\,,
\end{align}
and  the ${\cal O}(p^2)$ equation of state~\cite{Carignano:2016rvs} is 
\be
\epsilon(P) = -P + 2\sqrt{P(2f_\pi^2 m_\pi^2+P)}  \label{eq:eq-state}\,.
\ee
    
 \item The $K^+$ condensed phase is favored for
\begin{align}
\mu_S&>m_K-\frac12 \mu_I\,,\nonumber \\ 
\mu_S&>
\frac{-m_\pi^2+\sqrt{(m_\pi^2-\mu_I^2)^2+4 m_K^2 \mu_I^2 }}{2 \mu_I}\,,
\end{align}
resulting in the vacuum in Equation~\eqref{eq:kaon_condensate}
with 
\be
\cos \alpha_K=\left( \frac{m_K}{\mu_K}\right)^2\,, \ee 
where $\mu_K=\mu_I/2 +\mu_S$ is the relevant combination of chemical potentials, because $K^+$ has isospin $1/2$ and strangeness $1$.
The condensates are given by
\begin{align}
\sigma_u &= \sigma_s = \sigma_0 \cos  \alpha_{K}   \qquad \sigma_d = \sigma_0 \nonumber\\
\langle \pi^+ \rangle & =  0  \qquad \langle K^+ \rangle  =  \sigma_0 \sin  \alpha_{K}\,,  
 \end{align}
and the normalized pressure by
 \be
P = \frac{f_\pi^2\mu_K^2}{2} \left(1- \frac{m_K^2}{\mu_K^2}\right)^2 \,, \label{eq:pressure-pions}
\ee
which consents to obtain  the number densities 
\begin{align}
n_I =  \frac{f_\pi^2 \mu_K}2 \left(1-\frac{m_K^4}{\mu_K^4}\right)\qquad
n_s  = 2 n_I\,.
\end{align}
The EoS is 
\be
\epsilon(P) = -P + 2\sqrt{P(2f_\pi^2 m_K^2+P)}  \label{eq:eq-state}\,.
\ee
\end{itemize}

As we shall see in Section~\ref{sec:lattice}, these $\chi$PT results are in good agreement with the NJL and LQCD results close to the second order phase transitions. 

Let us briefly comment on  the difference between the charged meson condensed phases and the neutral kaon condensed phase. The expression of the various thermodynamic quantities in the neutral kaon condensed phase can be obtained from those of the charged kaon condensed phase by replacing $\mu_I \to - \mu_I$, due to the fact that $K_0$ has  isospin $-1/2$. The  relevant difference, as we have already noted is Section~\ref{sec:group}, is that the  $K_0$ and the $\bar K_0$  condensed phases are  superfluid, while the charged meson condensed phases are superconductors.   In the latter case one can  determine the screening masses of the electromagnetic field by gauging the $U(1)_Q$ subgroup of the chiral group, see Equation \eqref{eq:G_group},  resulting in  the Debye and Meissner screening masses~\cite{Mammarella:2015pxa}
\be\label{eq:screening}
M_D^2 = M_M^2 = f_\pi^2 e^2 (\sin\alpha)^2 \,,
\ee
where the Debye mass is related to the electric charge susceptibility, see for instance~\cite{bellac2000thermal}, while  
the nonvanishing value of the Meissner mass implies that  the system is a superconductor. 

\subsubsection{Low-Energy Excitations}\label{sec:low_energy}

Once the ground state has been identified, one can determine the low-energy fluctuations by an appropriate expansion. This is quite useful also because it allows us to identify the NGB. We briefly illustrate the procedure for the two-flavor case. A useful parameterization  is
\be\label{eq:single_sigma}
\Sigma = \cos \rho + i \hvf \cdot {\bm \sigma} \sin\rho \,,
\ee 
where the radial field, $\rho$,   and the unit vector field, $\hvf$, encode in a nontrivial way the three pion fields. By this parameterization the LO $\chi$PT  Lagrangian takes the form obtained in~\cite{Carignano:2016lxe} 
\begin{align}
\label{eq:lag-rho-phi}
{\cal L}=& \frac{f_\pi^2}{2} \left(\partial^\mu \rho\partial_\mu \rho + \sin^2\rho \; \partial^\mu \hf_i\partial_\mu\hf_i  -2 m_\pi \gamma \sin^2\rho  \; \epsilon_{3i k} \hf_i \partial_0 \hf_k \right) -V(\rho)\,, 
\end{align}
where
\be\label{eq:V}
V(\rho) = -f_\pi^2 m_\pi^2\left(\cos\rho + \frac{\gamma^2}{2}\sin^2\rho\right)\,,
\ee
is the potential and $\gamma = \mu_I/m_{\pi}$ is the control parameter. 
From the ground state analysis we know that the pion condensed phase is  favored for $|\gamma| >1$,
and the minimum of the potential is attained for the radial field vev, $\bar \rho = \alpha_\pi$, see Equation~\eqref{eq:alpha_pi}.

The low-energy radial and angular excitations  can now be introduced as follows~\cite{Mammarella:2015pxa, Carignano:2016lxe,Lepori:2019vec}
\be
\rho = \alpha_\pi + \chi\,, \qquad \hvf =(\varphi_1,\varphi_2)\,,
\ee
where we have neglected the fluctuation of the $\varphi_3 \equiv \pi_0$ field, because it decouples.  We can parameterize the angular field by    
\be\label{eq:def_hf}
\hvf= (\cos \theta,\sin \theta)\,,
\ee
where  $\theta$ is the  Bogolyubov mode, and rescaling  the radial  field as $\chi \to \chi/f_\pi$   and the Bogolyubov mode as  $\theta \to \theta/ f_\pi \sin\bar\rho$, one obtains the quadratic Lagrangian 
\be
{\cal L}= \frac{1}{2} \partial^\mu \chi\partial_\mu \chi -    \frac{1}{2} m_\chi^2 \chi^2 +  \frac{1}2  \partial_{\mu} \alpha \, \partial^{\mu} \alpha  -g \chi\partial_0\alpha\,,
\ee
where $m_\chi = m_\pi \gamma \sin\alpha_\pi$ is the mass of the radial field fluctuations and $g =  2 \mu_I $ is the  coupling
between the oscillations of the radial and the angular fields. The mass of the radial mode vanishes at the phase transition to the normal phase because it is a second order  phase transition. 
The Bogolyubov field seems to propagate at the speed of light, but integrating out the radial fluctuations one obtains the actual NGB with a phonon-like dispersion law  
\begin{align}\label{eq:disp_phonon}
E_\text{ph}&= c_s p\,,
\end{align}
thus propagating at the sound speed~\cite{Son:2000xc}, 
\begin{align}\label{eq:cs}
c_s & =\sqrt{\frac{\partial P}{\partial n_I}}=  \sqrt{\frac{\gamma^4-1}{\gamma^4+3}}\,.
\end{align}

Alternatively,  by diagonalizing the quadratic Lagrangian one obtains the dispersion laws
\begin{align}
E_\pm &= \sqrt{p^2+\frac{m_\text{eff}^2}{2}\pm  \sqrt{\left( \frac{m_\text{eff}^2}{2}\right)^2 + g^2 p^2}}\,,
\end{align}
where the low momentum expansion of the $E_-$ field coincides with the NGB dispersion law and the other mode with mass
\begin{align}
m_\text{eff}^2&= m^2_\chi+g^2=m_\pi^2\frac{\gamma^4+3}{\gamma^2}\,,
\end{align}
is  the rotated radial mode. In conclusion, the  low-energy modes correspond to a NGB with dispersion law  in Equation~\eqref{eq:disp_phonon} and to a radial mode with mass  $m_\text{eff}$.

\subsection{The Nambu-Jona Lasinio  Model}\label{sec:NJL}
The  meson condensed phases can also be studied by  a  modeling of the strong interaction by  contact interaction terms~\cite{Barducci:1990sv,Toublan:2003tt, Barducci:2004tt,Barducci:2004nc, He:2005nk, Ebert:2005cs, Ebert:2005wr,Mukherjee:2006hq, Andersen:2007qv,Abuki:2008tx, Abuki:2008wm, Xia:2014bla,Khunjua:2018jmn, Avancini:2019ego}, see~\cite{Khunjua:2019nnv}  for a brief recent review. These  models stem from the original work by  Nambu and Jona Lasinio~\cite{Nambu:1960xd,Nambu:1961tp,Nambu:1961fr} of a pre-QCD    Lagrangian for  the description of the strong interaction by   contact  interaction terms: 
\begin{align}\label{eq:NJL_lagrangian_original}
\mathcal{L}=\bar{\psi}\left[i\gamma^{\mu}\partial_\mu - M \right]\psi + G [(\bar{\psi} \psi)^2 + (\bar{\psi} i\gamma_5 \bm\sigma\psi)^2] \,,
\end{align}
where $\psi$ is the two-nucleon isodoublet, $M$ is the pertinent mass matrix and $G$ is a dimensional coupling.   The interaction  preserves the global symmetry group G, see Equation~\eqref{eq:G_group},  for a proper description of hadronic matter.  The nucleons emerge as quasiparticle states and the  spontaneously breaking of the chiral symmetry leads to the appearance of mesons. The model is   based on an analogy with the BCS theory of superconductivity describing the electron interaction by means of a local interaction term with no gauge fields.
It is not completely specified until a regularization scheme is provided and the value of the coupling constant is fixed. 

In the modern view~\cite{Ebert:1982pk,Ebert:1985kz, Klevansky:1992qe,Buballa:2003qv,Toublan:2003tt}, the model describes  quark matter  with an  effective  contact  interaction term that preserve the chiral symmetries of QCD. The spinor $\psi$ in Equation~\eqref{eq:NJL_lagrangian_original}  now represents the quark fields and $M$ the corresponding mass matrix, see Equation~\eqref{eq:quark_mass_matrix}. The NJL model (eventually supplemented by Polyakov loop terms) has been  applied to study the entire QCD  phase diagram in Figure~\ref{fig:phase_diagram}. The major phenomenological shortcoming of the quark NJL  model is that it  does not provide a confinement mechanism, indeed it has no gauge dynamics. Moreover, the presence of a dimension $6$ operator requires an ultra-violet   regularization scheme~\cite{Klevansky:1992qe}, which in the most used approximations is   a hard cutoff at the $\Lambda \sim 1$ GeV scale or a form factor of the form~\cite{Alford:1998mk}
\be\label{eq:form_factor}
F(\bm p^2) = \frac{\Lambda^2}{\bm p^2 + \Lambda^2}\,,
\ee
to mimic the asymptotic freedom property of QCD.  The coupling constant and the bare   quark masses are  then fixed to reproduce the low energy physics. Typical values of these quantities are  \be
G \Lambda^2 \simeq 6 \qquad m\simeq 1.5 \text{ MeV}  \qquad m_s\simeq 50 \text{ MeV} \,,
\ee
see however~\cite{Klevansky:1992qe,Buballa:2003qv}. 
The NJL model is a  useful tool for a qualitative and semiquantitative exploration of the properties of hadronic matter, however the obtained results  depend on the choice of these parameters and on the regularization scheme employed. Unfortunately, one cannot  systematically improve the model because no expansion parameter can be identified.  Despite these limitations, the NJL Lagrangian  is Lorentz  invariant, with the chiral symmetry   realized and spontaneously broken exactly as it is expected to happen in QCD: by a chiral condensate. Moreover,  the chiral symmetry can be  explicitly broken   by the inclusion of small current quark masses.

There is a certain degree of uncertainty in the form of the NJL Lagrangian. In the two-flavor case most of the authors retain the form in Equation~\eqref{eq:NJL_lagrangian_original}, although different chirally symmetric interactions can be written. This increases the number of phenomenological parameters that have to be fixed. Following~\cite{Asakawa:1989bq,Klevansky:1992qe, Buballa:2003qv}, the NJL Lagrangian can be generalized to
\be
{\cal L} =\bar{\psi}\left[i\gamma^{\mu}\partial_\mu - M \right]\psi  + {\cal L}_1+{\cal L}_2 \,,  
\ee
where    the two interaction terms are
\begin{align}
{\cal L}_1 &= G_1 [(\bar{\psi} \psi)^2 +(\bar{\psi} i\gamma_5 \bm\sigma\psi)^2 +(\bar{\psi} \bm\sigma\psi)^2+  (\bar{\psi} i\gamma_5 \psi)^2] \nonumber \\
{\cal L}_2 &= G_2   [(\bar{\psi} \psi)^2 +(\bar{\psi} i\gamma_5 \bm\sigma\psi)^2 -(\bar{\psi} \bm\sigma\psi)^2-  (\bar{\psi} i\gamma_5 \psi)^2] \,. 
\end{align}

The first term preserves  the $U(1)_A$  symmetry while the second  term explicitly breaks it. Whether~ or not the latter is comparable with the first depends on nonperturbative effects, indeed the $U(1)_A$ breaking term   is supposed to describe  the interaction mediated by some instanton configurations of the gauge fields. 
Considering different values of the coupling constants $G_1$ and $G_2$ implies  distinct values of the chiral condensates of different flavors and a  different phase diagram. This issue emerges also in the three-flavor case, where  the NJL Lagrangian takes a slightly different  form~\cite{Bernard:1987sg}, 
\begin{align}\label{eq:NJL_lagrangian}
\mathcal{L}=&\bar{\psi}\left[i\gamma^{\mu}\partial_\mu - M \right]\psi + G \sum_{a=0}^8 [(\bar{\psi} \lambda_a \psi)^2 + (\bar{\psi} i\gamma_5\lambda_a \psi)^2] - K [ \det\bar{\psi}(1+\gamma_5)\psi + \det\bar{\psi}(1-\gamma_5)\psi]\,,
\end{align}
 where now $\psi^T=(u,d,s)$ and $G$ and $K$ are the two coupling constants analogous of $G_1$ and $G_2$;  the~ determinant term  removes the $U(1)_A$ symmetry. 
Considering the $U(1)_A$ symmetric Lagrangian with $K=0$, the authors of~\cite{Barducci:2004tt} find the same transition hexagon (solid red line) reported in Figure~\ref{fig:phases}, however the up and down  chiral condensates split and at large values of the chemical potentials they find   phase transition lines not present in Figure~\ref{fig:phases}. 

In the following we will focus on the {\it traditional} two-flavor NJL model, with $G_1 =G_2 =G/2$ corresponding to the Lagrangian in Equation~\eqref{eq:NJL_lagrangian_original}, and thus maximally violated $U(1)_A$ symmetry. The~ presence of a medium can be  described by a covariant derivative analogous to Equation~\eqref{eq:covariant_quarks}, 
which~ takes into account  the baryon and isospin chemical potentials. With the NJL model one can explore  the entire QCD phase diagram (with the limitations discussed above), a clear advantage with respect to the $\chi$PT approach which can hardly investigate the effect of the baryon chemical potential. To obtain the properties of the vacuum and of the low-energy excitations one can perform a Hubbard-Stratanovich transformation introducing the collective boson variables
\be
\sigma_f(x) = - \frac{4 G}{\Lambda} (\bar\psi_f \psi_f)  \qquad \pi_{a} = - \frac{2 G}{\Lambda} (\bar\psi \gamma_5 i \lambda_a \psi + \text{h.c.})\,,
\ee
corresponding to scalar and pseudoscalar fields. Their expectation values are determined by minimizing the one loop effective potential, or equivalently, by solving the coupled gap equations. By this analysis it has been confirmed that the pion condensed phase  sets in at $\mu_I=m_\pi$, see \cite{He:2005nk}. Moreover, it has been determined the dependence of the condensates on the chemical potentials and on the temperature.

At vanishing temperature, in the two flavor case the grand potential has the particularly simple~ expression
\be
\Omega= G (\sigma^2 + \langle \pi \rangle^2) - \frac{3}{2 \pi^2}\int_0^\Lambda dk k^2 (E_+ +E_-)\,, 
\ee
where we have used the hard cutoff procedure and the quark quasiparticle dispersion laws are given~ by
\begin{align}
E_{q,\pm} = \sqrt{\left(E_k\pm \frac{\mu_I}{2}\right)^2 + 4 G^2 \langle \pi \rangle^2}\,,
\end{align}
where $E_k = \sqrt{k^2 + \bar m^2}$  with $\bar m=m - 2 G \sigma$ the effective quark mass. Therefore, the effect of the chiral condensate is a shift of the quark masses~\cite{Klevansky:1992qe}, while the pion condensate opens a gap between the quasiparticle dispersion laws. This is the typical effect of condensation on the quasiparticle spectrum, as it indicates the formation of correlated pairs of fermions. In the pion condensed phase it costs additional energy to produce quasiparticle fermionic  excitations because of quark-antiquark   pairs. Clearly, this picture of meson condensation cannot be directly compared with the $\chi$PT results of the previous section, because the considered degrees of freedom are different. However, as we shall discuss below, the values of the pion and chiral condensates  can be compared, as well as various thermodynamic quantities. Moreover, the three flavor NJL model phase diagram obtained in~\cite{He:2005nk} is in agreement with the  theory group expectation in Figure~\ref{fig:phases} and  quantitatively very similar to that obtained in $\chi$PT.  

In the NJL model it is possible to include an electron (or positron) background to neutralize the pion electric charge. When requiring the electrical neutrality~\cite{Ebert:2005cs, Ebert:2005wr}, the NJL models tend to disfavor the appearance of the pion condensed phase~\cite{Andersen:2007qv, Abuki:2008wm}. At the physical  point, corresponding to a neutral configuration in hydrostatic equilibrium,  the pions do not condense~\cite{Andersen:2007qv}. 

%

\subsection{Comparison with Lattice QCD }\label{sec:lattice}

The  LQCD simulations  are numerical implementation of the  QCD action on a discretized grid; the~ relevant physical results are then obtained performing the limit to the continuum. These~ simulations
 can lead to the precise determination of many hadronic quantities and can provide numerical evidence for conjectured properties of strongly interacting matter.  The  LQCD simulations have been very successfully used  for simulating hadronic matter  in vacuum, but dealing with in medium effects poses a series of  problems. The most important one is that the LQCD simulations at finite baryonic density   are hampered by the so-called sign problem. Very briefly, in the LQCD simulations with dynamical quarks  the Dirac degrees of freedom are typically integrated out, {see for instance~\cite{Smit:2002ug,Gattringer:2010zz}},  resulting in a partition function that can be written as the {euclidean path integral}
\be
{\cal Z} \sim \int d A_\mu \,\, e^{-S (A_\mu) }\det \Delta_D\,,
\ee
where $S$ is the {euclidean} action, $A_\mu$ are the gauge fields and $\det  \Delta_D$ is the  determinant of the Dirac operators. The  standard Monte Carlo simulations are based on importance sampling of the possible gauge configurations.  This procedure  works if $\det  \Delta_D >0$, that is with a real and positive Euclidean path integral  measure.  At   nonvanishing  baryonic density the LQCD numerical technique becomes problematic  because the Dirac determinant is complex.  Although continuous progress for facing this problem has been reported over the years, see for example~\cite{Muroya:2003qs, Schmidt:2006us, deForcrand:2010ys, Philipsen:2012nu, Aarts:2015tyj}, as of yet it  is not  a feasible tool for exploring the QCD phase diagram at large $\mu_B$ and, in particular, the transition from the confined phase to the CSC phase in Figure~\ref{fig:phase_diagram}. See however~\cite{Alford:1998sd,Lombardo:1999cz,Cea:2012ev, Nishida:2003fb}   for different LQCD approaches  to the region with nonvanishing baryonic and isospin chemical potentials.

Since it is hard to manage baryons in LQCD simulations, people decided to ignore baryons.   This~ poses the LQCD simulations outside the beta-equilibrated  sheet, as discussed in Section~\ref{sec:Introduction},  to~ explore a part  of the  QCD phase diagram where the  outcomes of the numerical simulations can be  compared  with different methods, in particular with the $\chi$PT and the NJL results. The key point is indeed that the  LQCD simulations  at nonvanishing isospin  chemical potential and zero baryonic density are  not affected by the sign problem~\cite{Alford:1998sd}. This does not mean that this direction is without obstacles: the realization of multi-hadron systems in LQCD is an extremely challenging problem, see~\cite{Detmold:2015jda} for a review. There is a  wealth of LQCD results for the  pion condensation  
\cite{Kogut:2002zg,Kogut:2004zg, Beane:2007es,Detmold:2008fn, Detmold:2012wc, Endrodi:2014lja,Brandt:2016zdy, Brandt:2018wkp} while there has been little progress on kaons~\cite{Detmold:2008yn,Detmold:2011kw}. As we will see, the LQCD results on pion condensation are reliable only for     $\mu_I \le 2 m_\pi$, but  these simulations are  steadily improving and becoming more accurate, even with physical quark masses and external magnetic fields~\cite{Endrodi:2014lja,Brandt:2016zdy, Brandt:2018wkp} 

The LQCD simulations can be performed in the canonical or in the grand canonical ensembles. In~ the grand canonical simulation one discretizes on a lattice  the actual QCD Lagrangian in Equation~\eqref{eq:QCD_lagrangian} with the isospin  chemical potential as external source.  In this approach   the strangeness density has to be zero, as the strangeness chemical potential makes the measure complex. Moreover,  the QCD Lagrangian  has to be supplemented with a pionic source~\cite{Kogut:2002tm,Kogut:2002zg}
\be\label{eq:L_lambda}
{\cal L}_\lambda = i \lambda \bar \psi \gamma_5 \sigma_2 \psi\,,
\ee
to trigger the breaking of the $U(1)_I$ symmetry and to stabilize the numerical simulations.  Since the $\lambda$ term explicitly breaks the $U(1)_I$ symmetry, in the pion condensed phase there is  a  pseudo-NGB with vanishing mass in the $\lambda \to 0$ limit. Therefore, the physical interesting results are  obtained doing both  the continuum and the $\lambda \to 0$ limits. The first quenched LQCD simulations~\cite{Kogut:2002tm} already reported the expected behavior of the condensates; these results were  soon improved considering $N_f =2$ dynamical quarks in~\cite{Kogut:2002zg}.  To obtain more precise results one has to consider that
both the pion  and the chiral condensates depend in a rather non-trivial way on $\lambda$. Moreover,   the first simulations employed a  large pion mass. Progress with respect to these aspects has been reported in~\cite{Brandt:2017oyy} where the  $\lambda \to 0$ limit has been tackled by a reweighing technique in simulations with  $2$ light flavors and a heavy strange quark  at the physical pion mass.

In Figure~\ref{fig:condensates_NJL}  we  compare the pion and chiral condensates obtained by the LO $\chi$PT, see Section~\ref{sec:chiral}, by the two-flavor NJL model, see Section~\ref{sec:NJL},  and by the LQCD simulations of~\cite{Brandt:2017oyy}. 
 \begin{figure}[H]
\centering
  \includegraphics[width=.55\textwidth]{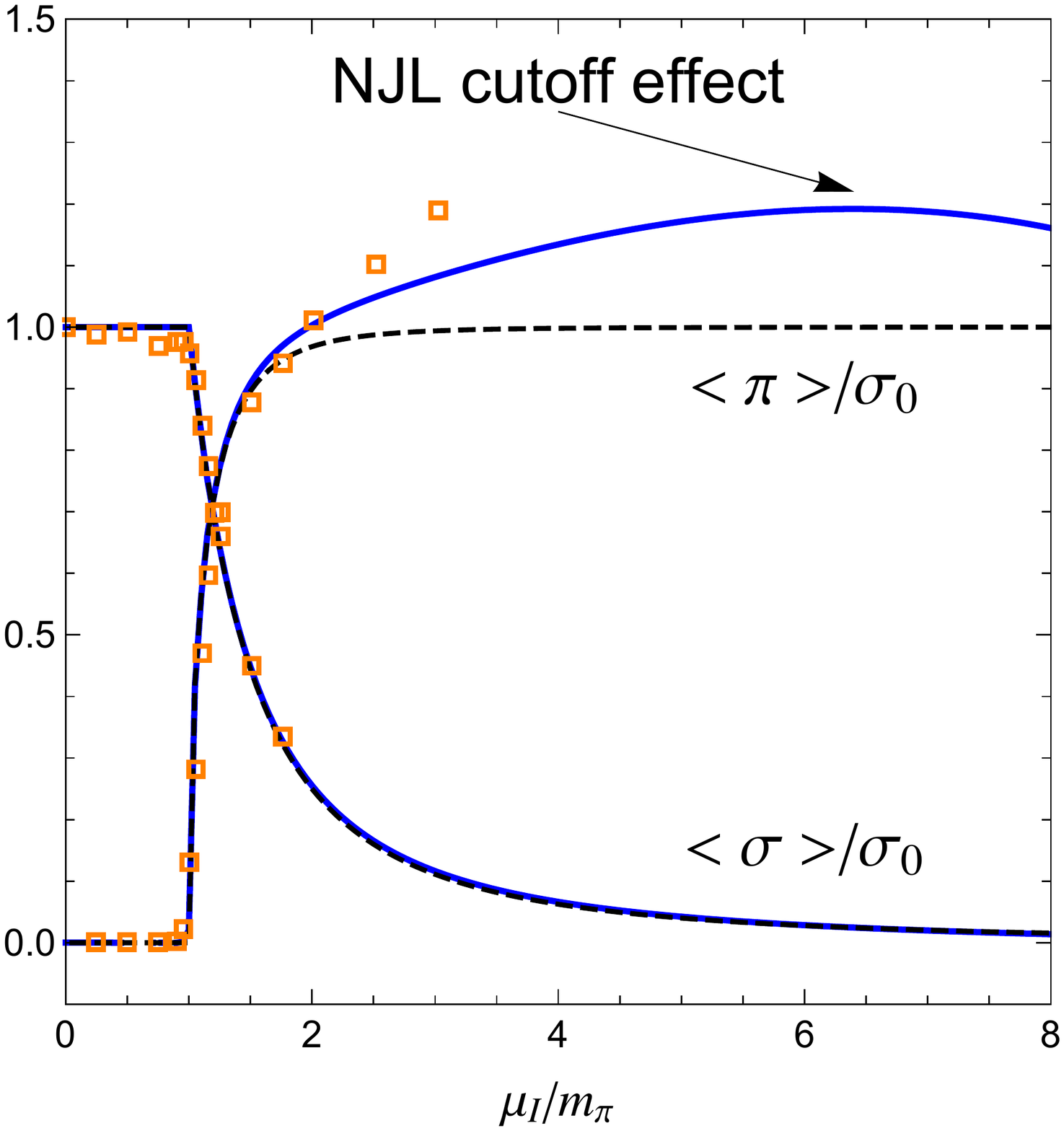}
\caption{Comparison of the chiral and pion  condensates obtained by $\chi$PT  (dashed black line),  by the two-flavor NJL model (solid blue line) and by LQCD simulations on a $6\times 24^3$ lattice by~\cite{Brandt:2017oyy} (orange squares).  The condensates have been normalized at the value of the scalar condensate in the normal phase (note that the LQCD data of~\cite{Brandt:2017oyy} have  been obtained at $T=113$ MeV), while the chemical potential is in units of the pion mass in the normal phase. Quite remarkably, the condensates obtained with the three methods overlap for $\mu_I \leq 2 m_\pi$. For larger values of the isospin chemical potentials the evaluation of the pion condensate becomes problematic: the NJL model feels the hard cutoff $\Lambda$; the $\chi$PT approaches the critical scale $\Lambda_\chi$; the LQCD simulations start to feel the lattice saturation effects~\cite{Kogut:2002zg}. }
\label{fig:condensates_NJL}
 \end{figure}
The chiral condensate obtained with the three approaches agree for any considered (or available) value of $\mu_I$. On~ the other hand, the pion condensates deviate at $\mu_I \sim 2 m_\pi$. In $\chi$PT the the two condensates obey the relation
\be
\sigma^2 + \langle \pi^+ \rangle^2 =\sigma_0^2\,,
\ee
 and therefore the pion condensate quickly saturates at $\mu_I \sim 2 m_\pi$. 
Both the NJL and the LQCD results for the pion condensate indicate that it exceeds $\sigma_0$  and it does not saturate at $\mu_I \sim 2 m_\pi$. This seems a robust result, although for larger values of $\mu_I$
 both the NJL and the LQCD approaches become problematic. 
The NJL results show a  non-monotonic behavior due to  the hard cutoff $\Lambda$, which  serves to mimic the asymptotic behavior of QCD, but that also signals the scale at which the NJL results are not under control. Similarly, the LQCD simulations feel the finite size lattice effects, indeed   for large value of $\mu_I$   saturation effects become important, see for example~\cite{Kogut:2002zg}. 
 Anyway, one can certainly regard this comparison as successful, in the sense that in the range  $\mu_I \lesssim 2 m_\pi$, where all the three  approaches are supposed to work, they give very similar results.
 
 We now turn to  the canonical approach~\cite{Detmold:2008fn,Detmold:2008yn,Detmold:2011kw, Detmold:2012wc}. The canonical LQCD simulations can explore quark matter at nonvanishing   isospin and strangeness density, while grand-canonical LQCD simulations can only deal with finite isospin density. In the canonical LQCD simulations  the isospin and strangeness density are fixed and  the corresponding chemical potentials are determined by thermodynamic relations. The description  of mesons in the canonical LQCD simulation is attained by the introduction of external sources with a fixed isospin or strangeness charge. In these simulations the calculation of the meson field correlator requires the computation of a  large number of Wick contraction of the quark fields on the lattice,  leading to time consuming and expensive calculations.   This is the  main limitation of the canonical lattice simulations.  Various  different algorithms for reducing theses costs have been developed in~\cite{Detmold:2012wc},
resulting in the simulation of  up to $72 \pi^+$ in configurations with spatial extents $L \sim 2, 2.5 $ and $3$ fm, resulting in isospin chemical potentials up to $4.5 m_\pi$~\cite{Detmold:2012wc}.
 
In Figure~\ref{fig:eps_vs_mu} we compare the energy density obtained with  the canonical LQCD simulations with  that of  the $\chi$PT and the NJL approaches. 
\begin{figure}[H]
\centering
\hspace*{-1.cm}  \includegraphics[width=.6\textwidth, height=.55\textwidth]{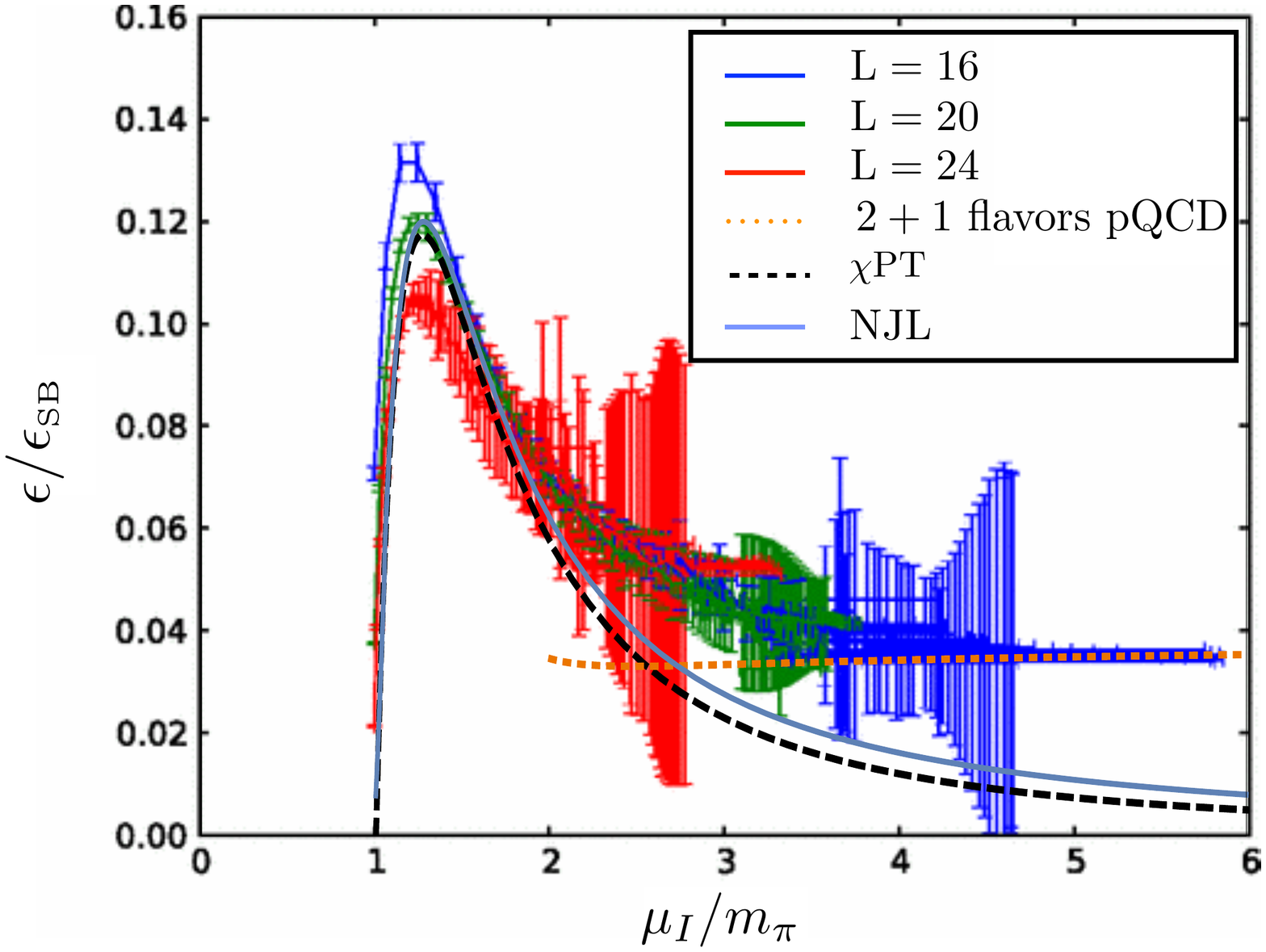}
\caption{Comparison of the energy density  over the Stefan-Boltzmann limit obtained by different methods. We report the lattice data points of the canonical simulations of~\cite{Detmold:2012wc} obtained at $T=20$ MeV with three  different lattice volumes. The pQCD results of~\cite{Graf:2015pyl}  (orange dotted line) indicate a constant asymptotic energy density. The $\chi$PT leading order results of~\cite{Carignano:2016rvs} (dashed black line) and the mean field NJL results  (solid blue line), see ~\cite{Xia:2013caa, He:2005nk}, perfectly reproduce the peak structure at  $\mu_I \simeq  1.27\,  m_\pi$. }
\label{fig:eps_vs_mu}
 \end{figure}
More precisely, in this figure it is shown the normalized energy density $\epsilon/\epsilon_\text{SB}$, where $ \epsilon_{SB}=9 \mu_I^4/(4 \pi^2) $ is the Stefan-Boltzmann limit, as a function of the normalized isospin chemical potential, $\mu_I/m_\pi$.  The dots with error bars are  the  results of~\cite{Detmold:2012wc} obtained with three different spatial volumes $L^3$. With increasing $\mu_I$ the error bars increase, signaling that lattice simulations  cease to be reliable at $\mu_I \sim 2 m_\pi$, as in the grand-canonical LQCD simulations discussed above. In this regime, the normalized energy  well agrees with the  $\chi$PT results (dashed black line), and~ with the NJL results (solid blue line). Both the  $\chi$PT and the NJL  curves perfectly capture the peak structure at low $\mu_I$, while they begin to depart from the LQCD  results at $\mu_I \sim 2 m_{\pi}$. The  $\chi$PT and the NJL   peak positions are  respectively at 
\begin{align}
\label{eq:mupeak_chi}
\mu_{I, \chi\text{PT}}^\text{peak} & = \({\sqrt{13} -2}\, \)^{1/2} m_\pi \simeq 1.27\, m_\pi \,,\\
\mu_{I, \text{NJL}}^\text{peak} & \simeq  1.27\,  m_\pi\,,
\end{align}
where the $\chi$PT results are  independent of $f_\pi$, while  the NJL results are not very sensitive to the parameter set used. The LQCD results  of~\cite{Detmold:2012wc} are peaked at   \begin{equation}\mu_{I,\text{LQCD}}^\text{peak} = \{1.20,1.25,1.275\} m_\pi \,, \ee  where the different values are obtained for lattice sides   $L=\{16,20,24\}$, respectively; the~   continuum-linearly-extrapolated peak is at $ \mu_{I,\text{LQCD}}^\text{peak} \simeq 1.30\,m_\pi$. Therefore, also the canonical LQCD simulations are in agreement with the  $\chi$PT and the NJL  results for $\mu_I \lesssim 2 m_\pi$.

\subsubsection*{Nonvanishing Temperature}
Given the successful comparison of the  $\chi$PT, NJL and LQCD approaches at $T=0$, one may expect a similar agreement  at small temperatures. As we will see, the agreement between the three methods at $T\neq 0$ is much worse.  Herein we  report on the investigation of the phase diagram at $T \neq 0$ and $\mu_I \neq 0$ comparing the results for   the transition lines at $\mu_B=0$ separating the pion condensed phase from the normal phase at low $T$, and  between the pion condensed phase and the quark-gluon plasma, at high $T$. We show in Figure~\ref{fig:T-muI} the results obtained with the different approaches.  The  LQCD simulations of~\cite{Brandt:2017oyy} indicate  that at $\mu_B=0$  there is  in the $\mu_I, T$ plane a  chiral crossover line (shaded blue region) joining the  points   $(0,160)$ to the (pseudo) tricritical point $(140,151)$ (orange dot with error bars).  At the tricritical point the chiral crossover line joins the second order phase transition line (shaded green region). The LQCD results for the second order phase transition  are almost insensitive to the temperature for $T \lesssim  150$ MeV, then the phase transition line  becomes strongly temperature dependent, with a sort of ``T-like'' phase diagram shape. The mean-field NJL second order phase transition (solid blue line)~\cite{He:2005nk} shows a behavior similar to that of the LQCD simulations  for $T \lesssim 100$ MeV, then for higher temperatures the NJL results show a more pronounced  temperature dependence.   Eventually,  the NJL critical curve saturates with a critical temperature that is not sensitive to the isospin chemical potential for $500 \text{ MeV }< \mu_I < 1$ GeV  (not shown in the figure).  The analytic $\chi$PT  temperature dependence of the second order phase transition has been obtained in~\cite{Splittorff:2002xn}
\begin{equation}\label{eq:muI_chi}
\mu_I(T)= m_\pi + \frac{1}{4 f_\pi^2}\sqrt{\frac{m_\pi^3 T^3}{2 \pi^3}}\zeta\left(\frac{3}2\right)\,,
\ee
and is reported in Figure~\ref{fig:T-muI} with a dashed black line. The $T^{3/2}$ behavior does not  agree with the LQCD nor with the NJL  results. The $\chi$PT results of~\cite{Loewe:2004mu} indicate an even stronger temperature dependence.  These results are somehow surprising, as one would expect $\chi$PT to work up to $T \lesssim 100$ MeV while  Figure~\ref{fig:T-muI} shows that  it is inconsistent with the LQCD low temperature behavior.

\begin{figure}[H]
\centering
\includegraphics[width=.6\textwidth]{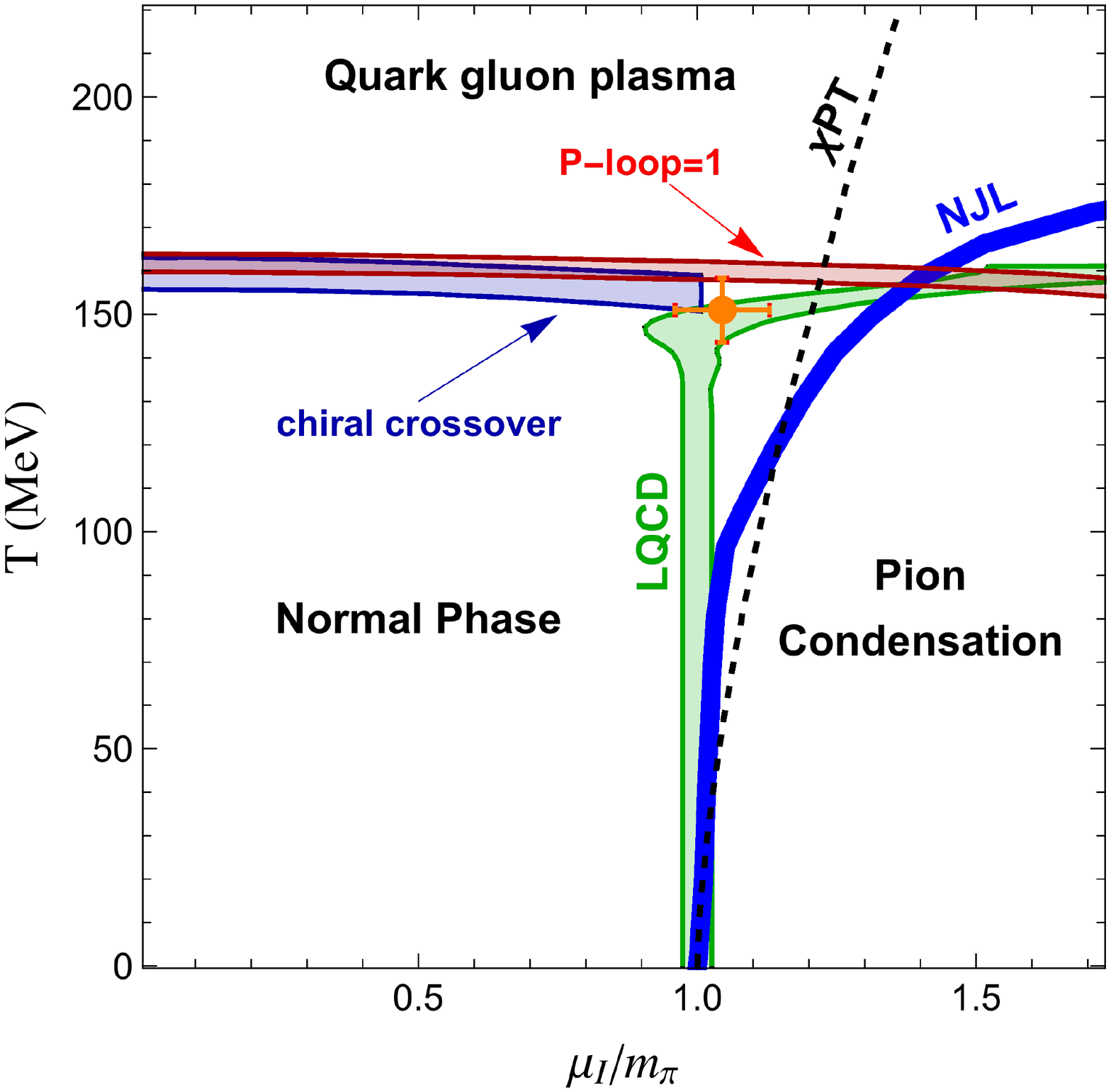}
\caption{Phase diagram of hadronic matter at $\mu_B=0$.   The green shaded area marked with LQCD corresponds to the second order phase transition separating  the pion condensed phase from  the normal phase, at low temperature,  and the chirally restored quark gluon plasma, at high temperature. These~ three phases meet at a (pseudo) tricritical point (orange dot with error bars), at $T_\text{tricritical} \simeq 151~ $MeV and $\mu_{I,\text{tricritical}} \simeq$ 140 MeV.  The shaded blue area marks the chiral crossover line which in part overlaps with a (probable) deconfinement phase transition (shaded red area), see the text for more details. These  results have been obtained in~\cite{Brandt:2017oyy}  by LQCD simulations. The NJL second order  transition  line (solid blue), see~\cite{He:2005nk}, overlaps with the LQCD data for $T \lesssim 100$ MeV. Then it shows a stronger temperature dependence.  The $\chi$PT second order transition line (dashed black), corresponding to Equation~\eqref{eq:muI_chi},  shows a  temperature dependence that disagrees with both the LQCD and NJL~ results.  }
\label{fig:T-muI}
 \end{figure}
Given the rather precise LQCD data, one should understand what are the origins of the discrepancies. The NJL results have been obtained by a hard cutoff scheme, maybe one can relax this requirement by a Pauli-Villars regularization scheme or by a form factor, as in~\eqref{eq:form_factor},  that does not completely eliminate the hard scale contribution. 
The improved $\chi$PT results of~\cite{Loewe:2004mu}  do not match the LQCD behavior at low $\mu_I$ but indicate a critical temperature that is independent of the chemical potential for $\mu_I >T$, which is in agreement with both the LQCD and NJL simulations. One should certainly try to understand what is  the $\chi$PT missing ingredient at lower temperatures. 

Quite remarkably, the LQCD simulations are now tackling the  color deconfinement transition as a function of the isospin chemical potential. Color deconfinement can be characterized by the behavior of the so-called Polyakov loop, see for instance~\cite{cheng1984gauge,Satz:2012zza},
\begin{equation}
P(r)= \frac{1}3 \Tr\, \Pi_{\tau =1}^{N_\tau} U_\tau(r)\,,
\ee
at large lattice spacing. The expectation value of the Polyakov loop is related to the correlation function between two static heavy quarks, therefore it is a measure of the strength of the color interaction. It~  has the important  property to vanish in the color confined phase of pure gauge theories~\cite{Satz:2012zza}.
A~ first inspection of the Polyakov loop dependence at nonvanishing $\mu_I$ and $T$  has been done in~\cite{Brandt:2017oyy}, by a $24^3\times 6$ lattice.  Various lines of constant values of the Polyakov loop have been obtained to infer the position of the deconfinement critical temperature.   In Figure~\ref{fig:T-muI} we report the line of~\cite{Brandt:2017oyy} (shaded red line) corresponding to  $P=1$, which in their notation can be taken as  indicating the color deconfinement transition.   
This  line partially overlaps with  the chiral crossover line, but then it starts to bend inside the pion condensed phase. These preliminary results should be tested with different lattice spacings. Quite interestingly,  the results of~\cite{Brandt:2017oyy} indicate that the  deconfinement line is quite insensitive to the presence of the pion condensate, or equivalently, to the melting of the chiral condensate at large $\mu_I$. We~ expect that for sufficiently large $\mu_I$  the BEC pion condensate turns  in a BCS condensate. In~ this case the deconfined quarks should  form quark-antiquark Cooper pairs, pretty much as in the CSC phase, but with an important difference: the BCS pairs in this case can be color singlets. Therefore, in~ this case  there is no need to have  color deconfinement nor any phase transition at all. Quite generally, indeed, there  is no phase transition between the BEC  and the BCS phases~\cite{giorgini-review}. Moreover, it would be interesting to see the behavior of the energy density of the system as a function of $T$ for a fixed $\mu_I$, as  we expect that the deconfinement phase transition should  induce a rapid increase of the energy density due to the liberation of the quark and gluon degrees of freedom.

\section{Conclusions}\label{sec:conclusions}

We have briefly reviewed the meson condensation phenomenon happening when the isospin or the strange chemical potentials exceed a critical value. We have clarified that it is unclear whether or not these phases can be  realized in Nature. In vacuum all mesons are unstable, therefore a stable meson can only exist in a dense medium, as in compact stars. In the core of these stellar objects the large number of electrons may stabilize the $\pi^-$,  however the problem is that with increasing density other particles  compete with $\pi^-$ to share the excess electron charge. By a simple noninteracting model we have seen how  the electron negative charge is drained off into  $\Sigma^-$ states favoring the strangeness production. The strong interactions can modify this picture, but it is unclear  whether they  favor or disfavor the appearance of stable pions. Quite recently, it has been proposed that pion stars consisting of pions and charged leptons may exist~\cite{Carignano:2016lxe,Brandt:2018bwq, Andersen:2018nzq}.  The astrophysical observation of this exotic  star would certainly be a smoking gun of a macroscopic coherent state of pions.

Although it is unclear whether the meson condensed phases are realized in compact stars or in any other physical setting, they are interesting by themselves. The reason is that they allow us to explore the properties of QCD in a regime in which various methods overlap. In particular, the~ $\chi$PT, the~ NJL and the LQCD approaches give similar results at vanishing temperature for $\mu_I \lesssim 2 m_\pi$. The~ $\mu_I -\mu_S$  phase diagram in Figure~\ref{fig:phases}  finds   $\chi$PT and  NJL in excellent  agreement, and the LQCD numerical results have confirmed that the phase transition between the normal phase and the pion condensation phase is of the second order. Unfortunately, the entire $\mu_I -\mu_S$ phase diagram has not been completely explored by canonical LQCD simulations; such simulations  could allow to figure out whether mixed phases are~ realized.

These findings allow us to improve the  first version of the QCD phase diagram shown in Figure~\ref{fig:phase_diagram}. That diagram was  based on naive arguments on the strong interaction. We  now draw  in Figure~\ref{fig:phase_diagram_final} the QCD phase diagram in which we have fed the acquired knowledge. 
\begin{figure}[H]
\centering
\hspace*{-1.2cm}\includegraphics[width=.65\textwidth]{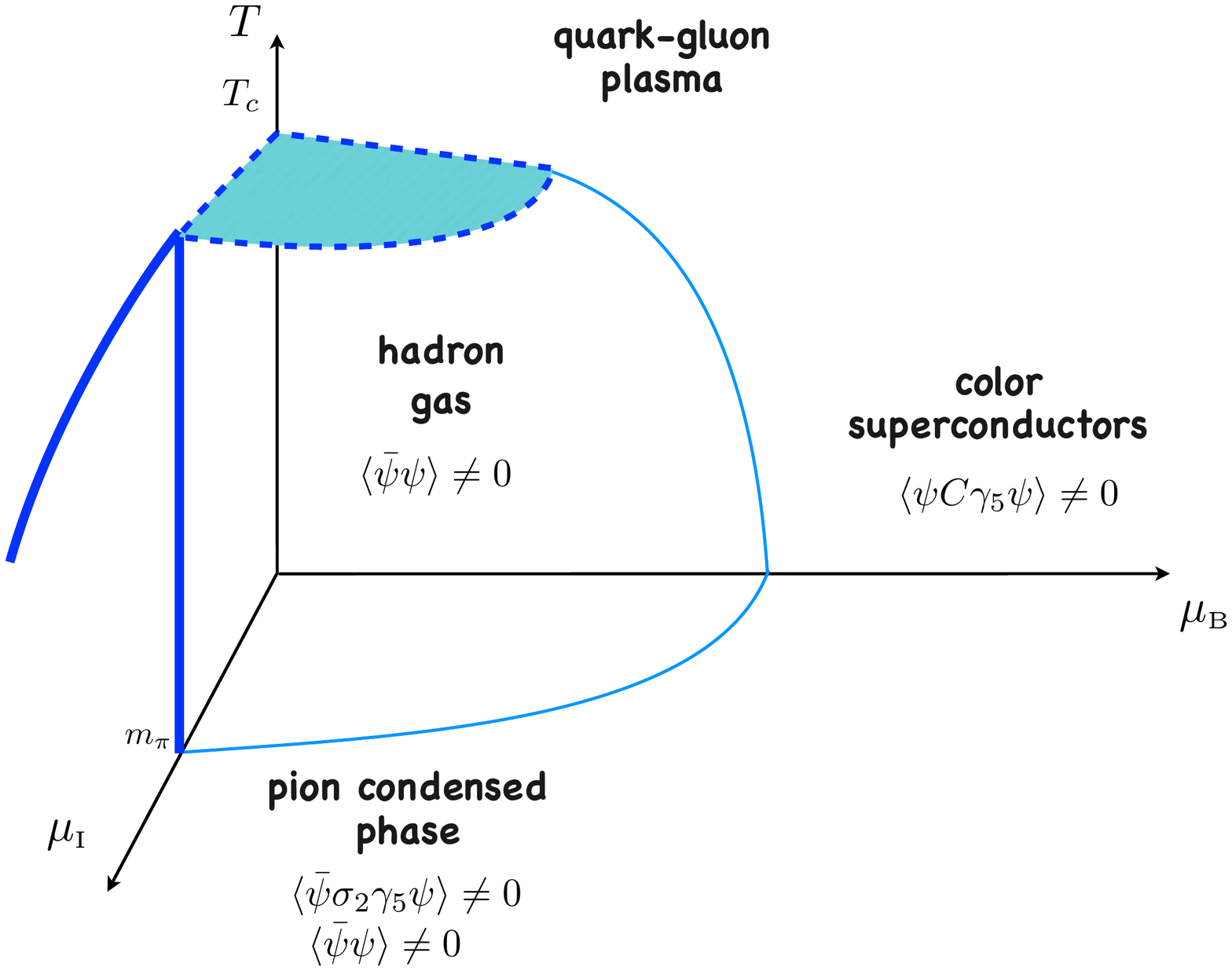}
\caption{Sketch of the phase diagram  of hadronic matter. The solid thick lines have been determined in the LQCD simulations of~\cite{Brandt:2017oyy}. For $\mu_B=0$ and $T \lesssim 150$ MeV, the transition to the pion condensed phase happens at $\mu_I=m_\pi$.  The shaded area on the top  corresponds to a chiral and deconfinement crossover. We have indicated with thin blue lines the possible transitions from the hadron gas phase to the color superconducting and to the pion condensed phases, although there are neither experimental data  nor LQCD simulations that support their existence. How the color superconducting phase turns in a quark-gluon plasma or in a pion condensed phase  is not known. }
\label{fig:phase_diagram_final}
 \end{figure}
The solid thick lines correspond to the second order phase transitions that are most tenable. The  LQCD simulations  give numerical support to the second order phase transition line at $\mu_I=m_\pi$, which should be  temperature independent up to $T\approx 150$ MeV, see also Figure~\ref{fig:T-muI}.  The dotted lines indicate  a chiral and (quite probably) deconfinement crossover.
Since there is a chiral crossover at small $\mu_B$ and $\mu_I=0$~\cite{Schmidt:2006us, deForcrand:2010ys, Philipsen:2012nu, Aarts:2015tyj} and since we have seen in the previous section that there is a chiral crossover at  $\mu_B=0$ and $\mu_I \lesssim m_\pi$, the most simple possibility is that there is an almost temperature independent chiral crossover region at  $T \simeq T_c$ (dashed area). There is indeed growing evidence that the chiral crossover extends in the  $\mu_B$, $\mu_I$ plane at an almost constant temperature $T\sim T_c$~\cite{Klein:2003fy,Toublan:2003tt, Barducci:2003un, Toublan:2004ks,He:2005nk}. 
The general result of these works is that the transition temperature smoothly decreases with   $\mu_I$ (or $\mu_S$), but the order of the phase transition is hard to establish. From Figure~\ref{fig:phase_diagram_final} it seems like  the hadron gas occupies a first octant sphere with the edges  cut by a vertical and an horizontal plane. 
However, there are still uncertain transitions, marked~ with a thin blue line. We have added a phase transition line between the hadron gas phase and the color superconducting phase, although there is no experimental data nor any LQCD simulation that supports it and may as well be a smooth crossover~\cite{Schafer:1998ef}, see also~\cite{Rajagopal:2000wf,Alford:2007xm}. We have also assumed that the hadron gas is separated from the pion condensed phase and/or the color superconducting phase by a transition (thin blue) line extending in the $T=0$ plane, which is just a guess.   
We have not shown in  Figure~\ref{fig:T-muI} any transition line between the color superconducting phase and the pion condensed phase. This   happens in a region where three different quark condensates compete and it is not at all obvious that the phase diagram has a simple form.

\vspace{6pt}

\funding{This research received no external funding. }

\acknowledgments{I would like to thank Gergely Endrodi and Bastian Brandt for sharing their LQCD data, Mark Alford and Gergely Endrodi for very useful suggestions and Jens Oluf Andersen for discussion. I thank the University of Bari and INFN for the support during the completion of this work.}

\conflictsofinterest{The author declares no conflict of interest.}

\reftitle{References}


\end{document}